\documentclass[a4paper,11pt]{article} 
  
\usepackage[english]{babel}
\usepackage[a4paper]{geometry}
\usepackage{amsmath}
\usepackage{amsthm}
\usepackage{amssymb}
\usepackage{bbm}
\usepackage{color}
\usepackage{enumerate}
\usepackage{authblk}

\usepackage{hyperref}         

\def\bR{\mathbb{R}}

\def\cC{\mathcal{C}}

\def\cV{\mathcal{V}}

\def\cF{\mathcal{F}}
\def\cG{\mathcal{G}}

\def\cN{\mathcal{N}}
\def\cE{\mathcal{E}}
\def\cK{\mathcal{K}}
\def\cH{\mathcal{H}}

\def\eps{\varepsilon}

\def\indic{\hbox{\raise-2pt \hbox{\indbf 1}}}

\let\==\equiv

\let\io=\infty
\let\0=\noindent

\def\*{{\hfill\break\null\hfill\break}}

\def\ie{\hbox{\it i.e.\ }}
\def\eg{\hbox{\it e.g.\ }}

\def\tende#1{\,\vtop{\ialign{##\crcr\rightarrowfill\crcr
             \noalign{\kern-1pt\nointerlineskip}
             \hskip3.pt${\scriptstyle #1}$\hskip3.pt\crcr}}\,}
\def\otto{\,{\kern-1.truept\leftarrow\kern-5.truept\to\kern-1.truept}\,}

\newtheorem{theorem}{Theorem}[section]  

\numberwithin{equation}{section}

\def\tl#1{{\tilde{#1}}}
\def\be{\begin{equation}}
\def\ee{\end{equation}}

\newcommand{\hc}{\mbox{h.c.}}

\let\a=\alpha \let\b=\beta    \let\g=\gamma     \let\d=\delta     \let\e=\varepsilon
        \let\k=\kappa     
                  \let\p=\pi        \let\r=\rho
\let\s=\sigma \let\t=\tau            
        
        \let\L=\Lambda    
             
\let\O=\Omega

\def\aa{\mathfrak{a}}
\def \blue#1 {\textcolor{blue}{#1}}
\def \red#1 {\textcolor{red}{#1}}

\definecolor{lightblue}{rgb}{0, 0.33, 0.71}

\title{A second order upper bound on the ground state energy of a Bose gas beyond the Gross-Pitaevskii regime}
\author{ Giulia Basti}
\affil{Gran Sasso Science Institute,\\Viale Francesco Crispi 7, 67100 L'Aquila, Italy}

\begin{document}

\maketitle

\begin{abstract}
We consider a system of $N$ bosons in a unitary box in the grand-canonical setting interacting through a potential with scattering length scaling as $N^{-1+\k},$ $\k\in (0,2/3).$ This regimes interpolate between the Gross-Pitaevskii regime ($\k=0$) and the thermodynamic limit ($\k=2/3$). In \cite{BCS}, as an intermediate step to prove an upper bound in agreement with the Lee-Huang-Yang formula in the thermodynamic limit, it is obtained a second order upper bound on the ground state energy for $\k<5/9.$ In this paper, thanks to a more careful analysis of the error terms, we extend the result in \cite{BCS} to $\k<7/12$.
\end{abstract}

\section{Introduction and main result}
It was predicted by Lee, Huang and Yang in \cite{LHY} (see also \cite{Bog}) that the ground state energy per unit volume of a dilute Bose gas satisfies
\be\label{eq:LHY}
	e(\r)=  4 \pi \aa \r^2  \left[ 1+ \frac{128}{15\sqrt \pi} (\r \aa^3)^{1/2}+ o( (\r \aa^3)^{1/2}) \right]
\ee
where $\rho$ denotes the particle density of the gas, $\aa$ the scattering length of the interaction potential and dilute refers to the fact that the mean interparticle distance is much larger than the scattering length, \ie $\r \aa^3\ll1.$ The expansion \eqref{eq:LHY} is known as Lee-Huang-Yang formula and its rigorous proof has been an open problem for a long time. In fact, while the leading term was already derived in \cite{D} as an upper bound for hard sphere interactions, the matching lower bound was obtained 40 years later in \cite{LY}. On the other hand, it was only with \cite{YY} (we also mention \cite{ESY} where the correct constant is recovered in the weak coupling limit) that the next to leading order was proved to be correct as an upper bound for regular potentials and later, with \cite{BCS}, for all potentials in $L^3$. Finally, in \cite{FS}, the Lee-Huang-Yang correction was established as a lower bound for all $L^1$ potentials and in the recent paper \cite{FSII} for a larger class of potentials including the hard sphere case. Note that in the latter case the matching upper bound is still missing.\\
We mention that the fermionic analogue of the expansion \eqref{eq:LHY} predicted in \cite{HY} has not yet been proved, see \cite{LSS} and \cite{FGHP} where the first two orders are derived (due to the Pauli principle an extra term of order $\r^{5/3}$ appears).\\
In this paper we will discuss and complement the result obtained in \cite{BCS}. There, as in \cite{YY}, the core of the proof is to build a grand canonical trial state in a box with periodic boundary conditions whose size is changing with $\rho,$ in particular the side length is assumed to be $\r^{-\g}$ for some $\g>1.$ Indeed, following a well known localization procedure (see \eg \cite{R}),  this trial state can then be easily modified to provide a trial state with the correct energy on a lager box with Dirichlet boundary conditions. The latter can finally be replicated to recover the thermodynamic box in the limit. Note that in the limit a grand canonical trial state can be proved to give an upper bound on the canonical ground state energy. 

Due to the strategy just described, the paper \cite{BCS} produces a side result of some interest on its own. Namely, it provides an upper bound  correct up to the second order on the energy of an Hamiltonian acting on the Fock space built on a box whose side length is of the form $\r^{-\g}$ for some $\g>1$. By scaling the described setting is equivalent to consider $N$ bosons in the unitary box $\Lambda=[-1/2,1/2]^3\subset\mathbb{R}^3$ interacting through the Hamiltonian $\cH_N$ acting on the bosonic Fock space $\cF(\L)$ whose action on the $n$-particle sector is given by
\begin{equation}\label{eq:HN}
	\cH_N^{(n)} = \sum_{j=1}^n -\Delta_{x_j} + \sum_{1 \leq i,j \leq n} N^{2-2\k} V (N^{1-\k} (x_i - x_j))
\end{equation}
with $\k\in (1/2,2/3)$ (note that the request $k>1/2$ comes from the assumption $\g>1$ needed to use the localization technique but it is never used in the proof of \eqref{eq:E_N} below which remains valid for $0<\k<1/2$). In \cite{BCS} it has been shown that, under suitable assumptions on the potential $V,$ the ground state energy $E_N$ of the hamiltonian $\cH_N$, satisfies
\be\label{eq:E_N}
	E_N  \leq   4 \pi \aa N^{1+\k}  \bigg(1+\frac{128}{15\sqrt{\pi}}(\aa^3N^{3\k-2})^{1/2}\bigg)+  C N^{5\k/2}   \max \{ N^{-\e} \hspace{-.1cm} , N^{9\k-5 +6\eps}  \hspace{-.1cm} ,N^{21\k/4-3+3\e}\} 
\ee
for all $\k\in (1/2,2/3)$ and $\e$ such that $3\k-2+4\e<0.$ Let us stress that Eq. \eqref{eq:E_N}, whenever $\k<5/9,$ is just the equivalent of Eq. \eqref{eq:LHY} written for the rescaled hamiltonian \eqref{eq:HN} (note that the scattering length of the rescaled potential is given by $\aa/N^{1-\k}$ with $\aa$ the scattering length of the original potential).

The main idea to construct the trial state leading to \eqref{eq:E_N} is to first generate the condensate, since Bose Einstein condensation is expected to hold in the ground state of \eqref{eq:HN}, and then to add correlations acting with a Bogoliubov transformation. However, it is known (see \cite{ESY,NRS}) that a quadratic operator in not enough to capture the correct second order of the energy. In \cite{YY} the exponential of the sum of a quadratic and a cubic operator was taken into account to better describe correlations. On the other hand, in \cite{BCS} the exponential of a quadratic and of a cubic operator act separately, inspired by the methods developed in recent years in \cite{BBCS,BBCSacta} to study the Gross-Pitaevskii regime (corresponding to $\k=0$ in \eqref{eq:HN}). The drawback in considering the exponential of a cubic operator is the lack of explicit formulas for its action that makes computations harder. In particular, to handle the desired more singular regimes $\k>1/2$ new ideas are needed w.r.t. those used in \cite{BBCS,BBCSacta}. In fact, in \cite{BCS} the cubic operator is implemented as a non unitary operator acting directly on the vacuum with some crucial restrictions on the allowed momenta. 

 In \cite{BCS} it was stated as a remark that the same method could have been pushed to cover all $\k<7/12$ but such extension was out of the scope of that paper. Our aim here is to prove this statement thanks to a more careful analysis of some error terms. More precisely, we want to prove the following theorem.
\begin{theorem}\label{thm:main}
 Let $0<\k<7/12$ and $\eps > 0$ small enough. Let $V \in L^{3}(\bR^3)$ be non-negative, radially symmetric, with $\mathrm{supp}(V)\subset B_R(0)$ and scattering length $\aa$. Then, for all  $N$ large enough
\begin{equation}\label{eq:main}
E_N  \leq   4 \pi \aa N^{1+\k}  \bigg(1+\frac{128}{15\sqrt{\pi}}(\aa^3N^{3\k-2})^{1/2}\bigg)+ {C N^{5\k/2}N^{-\e}}.
\end{equation} 
\end{theorem}
We conclude this section with some comments about the scaling in \eqref{eq:HN}. As we already mentioned, for $\k=0$ one recovers the well known Gross-Pitaevskii regime. In this setting the expansion of ground state energy has been established to first order in \cite{LSY,LY,NRS} while the second order was proved in \cite{BBCSacta} (where also the low energy spectrum is derived) for all potentials in $L^3$ (and can be extended to all $L^1$ interactions as discussed in \cite{NT}). Recently, in \cite{BCOPS} (see also \cite{BCOPSII}) the second order correction has been established as an upper bound in the hard core case. On the other hand, regimes with positive $\k$ are considered in \cite{BrCS} where the expansion of the ground state energy up to second order is obtained for sufficiently small $\k$. We also mention \cite{ABS} and \cite{F} where Bose Einstein condensation is proved for $\k<1/43$ and $\k<2/5$ respectively (see also \cite{DG} where a similar but simpler regime is considered). Proving condensation for $\k=2/3$, \ie directly in the thermodynamic limit, is a challenging and widely open problem so far, see \cite{Be,Ba} for preliminary results. Note that all the mentioned results are valid in the canonical setting while, on the contrary, the grand canonical setting is considered here.

{\it Acknowledgment} The author gratefully acknowledge the support from the GNFM Gruppo Nazionale per la Fisica Matematica - INDAM.

\section{Definition of the trial state}
To get an upper bound on the ground state energy of the operator $\cH_N$ defined in \eqref{eq:HN} we have to exhibit a trial state whose energy is bounded by the r.h.s. of \eqref{eq:main}. We first rewrite the Hamiltonian using the bosonic creation and annihilation operators $a^*_p,a_p,$ $p\in \L^*=2\pi\mathbb{Z}^3:$
\[
	\cH_N=\sum_{p\in\L^*}p^2a_p^*a_p+\frac{1}{2N^{1-\k}} \sum_{p,q,r\in \L^*} \widehat{V} (r/N^{1-\k}) \, a_{p+r}^*a_q^*a_{q+r}a_{p}.
\]
Our trial state is defined as in \cite{BCS}, we recall here the definition referring the reader to \cite{BCS} for more details. First, we introduce the Weyl operator 
\[
W_{N_0}=\exp \big[ \sqrt{N_0} a_0^* - \sqrt{N_0}a_0 \big]
\]
where $N_0>0$ is a parameter to be fixed. The role of $W_{N_0}$ is to generate the condensate, indeed we expect most of the particle to be in the condensate wave function (\ie $\varphi(x)=1$ which is the ground state of the non interacting problem). Next, we have to take into account correlations among particles due to the presence of interaction. To this end we consider the solution to the Neumann problem on the ball $|x|<N^{1-\k}\ell:$
\[
	 \left[ -\Delta + \frac{1}{2} V \right] f_{\ell} = \lambda_{\ell} f_{\ell}
\]
with $0<\ell<1/2$ and with the boundary condition $f(x)=1$ if $|x|=N^{1-\k}\ell.$ Furthermore, we define $f_{N,\ell}(x)=f_{\ell}(N^{1-\k}x)$ and we set $\eta_p=-N\widehat{\omega}_{N,\ell}(p)$ where $w_{N,\ell}=1-f_{N,\ell}$ and the hat denotes the Fourier transform.
 Note that 
 $|\eta_p|\leq CN^\k p^{-2}$.
 To define the quadratic transformation we also have to introduce two sets of momenta: the set of low momenta
  $ P_L=\{p\in \L^*:|p|\leq N^{\k/2+\e}\}\subset\L^*$ and its complement $P_L^C.$
Then, to get the desired upper bound we will consider a Bogoliubov transform whose kernel coincides with $\eta$ on $P_L^C.$ On the other hand, on the set of low momenta we consider the kernel $\t$ defined by
\[
	\tanh(2\t_p) = - \frac{8 \pi \aa N^\k}{p^2 +8 \pi \aa N^\k}\,.
\]
We are now ready to introduce the Bogoliubov transformation
\[
	T_\nu=  \exp \bigg(\;\frac 12 \sum_{p \in \L^*_+} \nu_p \big(a^*_p a^*_{-p} - \hc \big) \;\bigg)
\]
where the coefficients $\nu_p$ are defined as follows: $\nu_p=\eta_p$ for $p\in P_L^{C}$ and $\nu_p=\tau_p$ for $p\in P_L$.
However, $T_\nu$ is still not enough to get the energy correct up to the second order and to give a more precise description of correlations we are going to consider a cubic operator. To do so we first introduce the notations $\g_p=\cosh(\nu_p),\,\s_p=\sinh(\nu_p)$. 
We also need two new sets of momenta: $P_H=\{p\in \L^*:|p|>N^{1-\k-\e}\}$ and $P_S=\{p\in \L^*: N^{\k/2-\e}\leq |p|\leq N^{k/2+e}\}\subset P_L.$ Let us mention that, considering the restriction of $\eta$ to $P_H$ (denoted by $\eta_H$) and the restriction of $\g,\s$ to $P_S$ (denoted by $\g_S,\s_S$ respectively) we have
\be\label{eq:eta_H}
	\| \eta_H \|^2 \leq C  N^{3\k-1 +\eps} \,, \quad \| \eta_H \|^2_{H^1} \leq C N^{1+\k} \,, \quad \|\eta_H \|_\io \leq C N^{3\k-2 +2\eps}  
\ee
and
\be\label{eq:tau_S}
	\begin{aligned}
	&\| \s_S\|^2 \leq C N^{3\k/2}\,, \qquad & \| \s_S\|^2_{H^1} \leq C N^{5\k/2+\eps}\,, \\
	&\| \g_S\s_S \|_1 \leq C N^{3\k/2 +\eps}, &\| \gamma_S \|_\infty^2, \| \sigma_S \|^2_\infty \leq C N^\eps.
	\end{aligned}
\ee
With this notation at hand we can finally define the desired cubic operator. Namely,
\be \label{eq:defA}
	\begin{aligned}
		A_\nu  =  \frac 1 {\sqrt {N}} \sum_{\substack{r\in P_H,v \in P_S:\\r+v\in P_H}} \eta_r \s_v \,a^*_{r+v} a^*_{-r} a^*_{-v} \Theta_{r,v}
	\end{aligned}
\ee
where the operator $\Theta_{r,v},$ for $r\in P_H, v\in P_S$ with $r+v\in P_S$ is defined by
\[\label{eq:defTheta} \begin{split}
\Theta_{r,v}= &\;  \!\!\prod_{s \in P_H}\!\Big[ 1-\chi(\cN_s>0)\chi(\cN_{-s+v}>0)\Big] \\ & \hspace{3cm} \times \prod_{w \in P_S}\! \Big[ 1-\chi(\cN_w>0)\chi(\cN_{r-w}+\cN_{-r-v-w}>0) \Big].
\end{split}\]
In \eqref{eq:defTheta} $\cN_p=a^*_pa_p$ counts the number of particles with momentum $p$ and $\chi$ is the characteristic function. 
The trial state we are going to consider to prove \eqref{eq:main} is obtained acting on the vacuum vector $\Omega\in \cF$ first with the operator $e^{A_\nu}$, then with $T_\nu$ followed by $W_{N_0}$; finally, since $e^{A_\nu}$ is not a unitary operator we have to normalize:
\[
	\Psi_N=\frac{W_{N_0}T_\nu e^{A_\nu}\Omega}{\|e^{A_{\nu}}\Omega\|^2}.
\]
Here, we fixed $N_0=N-\|\s_L\|^2.$\\
Let us stress that the role of the operator $\Theta_{r,v}$ in the definition of $A_\nu$ in \eqref{eq:defA} is to avoid certain relations among momenta created by the action of $e^{A_\nu}$ on the vacuum and this results in a drastic simplification of the computations. As discussed in \cite[Sec. 2]{BCS}, one can write
\[
	\begin{aligned}
		A_\nu^m\O = &\frac1{N^{m/2}}\sum_{\substack{r_1\in P_H,v_1\in P_S:\\r_1+v_1\in P_H}}\dots\sum_{\substack{r_m\in P_H,v_m\in P_S:\\r_m+v_m\in P_H}}\prod_{i=1}^m \eta_{r_i}\s_{v_i}\\
	 &\hspace{2cm} \times \theta(\{r_j,v_j\}_{j=1}^m)a^*_{r_m+v_m}a^*_{-r_m}a^*_{-v_m}\dots a^*_{r_1+v_1}a^*_{-r_1}a^*_{v_1}\O
	 \end{aligned}
\] 
where $\theta$ encodes all the restrictions mentioned above:
\[
\theta\big( \{r_j, v_j \}_{j=1}^{m} \big) = \prod_{\substack{i,j,k=1\\j\neq k}}^m\prod_{\substack{p_i\in\{-r_i,r_i+v_i\}\\p_k\in\{-r_k,r_k+v_k\}}}\d_{-p_i+v_j\neq p_k}.
\]
Then, setting $\xi_\nu=e^{A_\nu}\Omega$, one has
\begin{equation}\label{eq:norm_xi}
	\begin{aligned} 
 		\| \xi_\nu\|^2 &= \sum_{m\geq 0}\frac1{2^m m!}\frac1{N^{m}} \sum_{\substack{ v_1 \in P_S, r_1 \in P_H : \\  r_1 +v_1 \in P_H  } } \cdots \sum_{\substack{ v_m \in P_S, r_m \in P_H : \\  r_m +v_m \in P_H  } } \theta\big(\{r_j, v_j\}_{j=1}^{m}\big)  \, \prod_{i=1}^{m} \big( \eta_{r_i} +\eta_{r_i+v_i} \big)^2  \s_{v_i}^2 .
	\end{aligned} 
	\end{equation}

\section{Proof of the main theorem}
In this section we discuss the modifications which are needed to extend the result obtained in \cite{BCS} to a larger set of choices of $\k,$ proving Theorem \ref{thm:main}.\\
Let
\[
	\cG_N = T^*_\nu W^*_{N_0} \cH_N W_{N_0} T_\nu
\]
so that $\langle\Psi_N,\cH_N\Psi_N\rangle=\langle\xi_N,\cG_N\xi_N\rangle/\|\xi_N\|^2$. 
Moreover, let us introduce the kinetic energy operator $\cK=\sum_{p\in \L^*}p^2a^*_pa_p$ and the operators
\[
\cV_N^{(H)}= \frac{1}{2 N}\sum_{\substack{r\in \L^*,\, p,q \in P_H:\\ p+r,q+r\in P_H}}N^\k \widehat{V}(r/N^{1-\k}) a_{p+r}^*a_q^*a_{p}a_{q+r} 
\]
and
\[
\cC_N =\; \frac{\sqrt{N_0}}{ N} \sum_{\substack{p,r \in P_H\\ p+r \in P_S}} N^\k \widehat V(r/N^{1-\k})\, \sigma_{p+r} \g_p\g_r  \,(a^*_{p+r} a^*_{-p} a^*_{-r}  + \hc )\,. 
\]
Finally, let
\[
	\begin{aligned}
		C_{\cG_N} =\; &  \frac{N^{1+\k}}{2}  \widehat{V}(0) +\sum_{p\in {\L}^*_+}p^2\s_p^2 + \sum_{p\in \L^*_+} N^\k \widehat{V}(p/N^{1-\k})\s_p\g_p\\
&+ \sum_{p\in P_L}N^\k \widehat{V}(p/N^{1-\k}) \s_p^2   +\frac{1}{2N}\sum_{\substack{p,r\in \L^*_+\\r\neq p}} N^\k \widehat{V}(r/N^{1-\k})\s_p\s_{p-r}\g_p\g_{p-r} \\
&  -  \frac 1 N \sum_{v \in P_L} \s^2_v \sum_{p \in P_L^c}  N^\k \widehat V(p/N^{1-\k}) \eta_p \,.
	\end{aligned}
\]
It has been shown in
 \cite[Prop. 3.1]{BCS} that, for any $0<\k<2/3$ and $\e>0$ such that $3\k-2+4\e<0,$
\begin{equation}\label{eq:<xiGxi>}
		\langle\Psi_N,\cH_N\Psi_N\rangle= \frac{\langle \xi_\nu, \cG_N \xi_\nu \rangle}{\| \xi_\nu \|^2} \leq C_{\cG_N} + \frac{\langle \xi_\nu, (\cK+\cV_N^{(H)}  +\cC_N) \xi_\nu\rangle}{\| \xi_\nu \|^2} +\frac{\langle \xi_\nu,\cE\xi_\nu\rangle}{\|\xi_\nu\|^2}
\end{equation} 
 with 
 \be\label{eq:cE}
 	\frac{\langle \xi_\nu,\cE\xi_\nu\rangle}{\|\xi_\nu\|^2}\leq C N^{5\k/2} \cdot  \max \{ N^{-\eps}, N^{9\k - 5 + 6\e} , N^{21 \kappa /4 - 3 + 3 \eps } \}.
\ee
Moreover, $C_{\cG_N}$ and the expectation on $\xi_\nu$ of the operators $\cK,\cC,\cV_N^{(H)}$ provide  the correct energy up to an error which is small under the previous assumptions (see \cite[Sec. 3 and 5]{BCS}):
\be\label{eq:en}
	\begin{aligned}
	C_{\cG_N} + \frac{\langle \xi_\nu, (\cK+\cV_N^{(H)}  +\cC_N) \xi_\nu\rangle}{\| \xi_\nu \|^2}\leq 4 \pi \aa N^{1+\k} & \bigg(1 + \frac{128}{15 \sqrt{\pi}} \, (\mathfrak{a}^3 N^{3\kappa-2})^{1/2} \bigg)\\
	&+  C N^{5\k/2}  \max\{N^{-\e},N^{12\k-7+5\e}\}\,.
	\end{aligned}
\ee
The bound \eqref{eq:cE} was obtained in \cite{BCS} using suitable bounds on the expectation over $\xi_\nu$ of products of the kinetic energy $\cK$ and powers of the particle number operator $\cN=\sum_{p\in \L^*}a^*_pa_p.$ However, it is clearly not compatible with the claim that \eqref{eq:main} is satisfied for all $\k<7/12$ and is indeed responsible for the request $\k<5/9$ in \cite{BCS}. Therefore, in order to prove Theorem \ref{thm:main}, we have to obtain an improved estimate on the expectation of the error term $\cE$ coming from the quadratic transformation.

\medskip

{\it Remark} Note that the error of the form $N^{5\k/2}N^{12\k-7+5\e}$ appearing in \eqref{eq:en} and coming from the action of the cubic operator $e^{A_\nu},$ in particular from the restrictions on the allowed momenta encoded in the operator $\Theta$, cannot be improved with the methods presented here. Therefore, to treat $k>7/12$ new ideas are needed.

\medskip

 To improve the estimate \eqref{eq:cE} we first identify, with a careful reading of the proof of Prop. 3.1 in \cite{BCS}, those terms in $\cE$ giving the worst rate\footnote{The terms in the proof of \cite[Prop.3.1]{BCS} responsable for the worst rate come from the conjugation of the cubic and quartic terms. More precisely (using the notation used in \cite[Sec. 4]{BCS}), they are $F_2,F_3,$ the first two terms in $G_2,$ the first term in $G_3$ and $G_1-\cV_N^{(H)}.$} rewriting
\be\label{eq:cEsplit}
	\cE=\cE_1+\cE_2
\ee
with 
\be\label{eq:errok}
	\frac{\langle\xi_\nu\cE_1\xi_\nu\rangle}{\|\xi_\nu\|^{2}}
	\leq CN^{5\k/2-\e}
\ee
for all $\k<2/3$ choosing $\e$ small enough.
Thus, we can focus on $\cE_2$ which can be splitted as
\be\label{eq:cE2}
	\cE_2=\cE_{C}+\cE_{S}+\cE_{H}+\cE_{M}.
\ee
Here, $\cE_C$ is the cubic operator defined by
\[
	\begin{aligned}
		\cE_C=&\frac{\sqrt{N_0}}{N}\!\!\sum_{\substack{p\in P_H,r\in P_S:\\p+r\in P_H}}\!\!\! \a(p,r)(a^*_{p+r}a^*_{-p}a^*_{-r}+\hc)
	\end{aligned}
\]
where we introduced the notation
\[
	\a(p,r)=N^\k\bigg(\widehat V\Big(\frac r{N^{1-\k}}\Big)+\widehat V\Big(\frac p{N^{1-\k}}\Big)\bigg)(\g_r\g_p\s_{p+r}+\s_r\s_p\g_{p+r}).
\]
On the other hand $\cE_S,\cE_H$ and $\cE_M$ are quartic operators. In particular, in $\cE_H$ only operators with high momenta
appear and it is defined by
 \[
	\begin{aligned}
		&\cE_{H}=\cE_{H,1}+\cE_{H,2}\\
		&=\!\frac1{2N}\!\sum_{r\in \L^*}\!\!\sum_{\substack{p,q\in P_H:\\p+r,q+r\in P_H}}\hspace{-0.5cm}\b_1(p,q,r)a^*_pa^*_{q+r}a_qa_{p+r}+\frac1N\!\sum_{r\in \L^*_+}\!\!\sum_{\substack{p,q\in P_H:\\p+r,q+r\in P_H}}\hspace{-0.5cm}\b_2(p,q,r)a^*_{p+r}a^*_{-p}a_{q+r}a_{-q}
	\end{aligned}
\]
where
\[
	\begin{aligned}
		\b_1(p,q,r)=&N^\k\widehat{V}\Big(\frac r{N^{1-\k}}\Big)\big[(\g_p\g_q\g_{p+r}\g_{q+r}-1)+\s_p\s_{p+r}\s_q\s_{q+r}+\g_p\g_{p+r}\s_q\s_{q+r}\big]\\
		\b_2(p,q,r)=&N^\k\widehat{V}\Big(\frac r{N^{1-\k}}\Big)\g_{p+r}\g_{q+r}\s_p\s_q.
	\end{aligned}
\]
Conversely, in $\cE_S$ only momenta in $P_S$ are involved. In fact, setting
\[
	\begin{aligned}
	\zeta_1(p,q,r)=&N^\k\widehat{V}\Big(\frac r{N^{1-\k}}\Big)(\s_p\s_{p+r}\s_q\s_{q+r}+\g_p\g_q\g_{p+r}\g_{q+r}+\g_p\g_{p+r}\s_q\s_{q+r})\\
	\zeta_2(p,q,r)=&N^\k \widehat V\Big(\frac r{N^{1-\k}}\Big)\g_{p+r}\g_{q+r}\s_p\s_q.
	\end{aligned}
\]
we have
\[
	\begin{aligned}
		&\cE_{S}=\cE_{S,1}+\cE_{S,2}\\
		&=\!\frac1{2N}\!\sum_{r\in \L^*}\!\!\sum_{\substack{p,q\in P_S:\\p+r,q+r\in P_S}}\hspace{-0.5cm}\zeta_1(p,q,r)a^*_pa^*_{q+r}a_qa_{p+r}+\frac1N\!\sum_{r\in \L^*_+}\!\!\sum_{\substack{p,q\in P_S:\\p+r,q+r\in P_S}}\hspace{-0.5cm}\zeta_2(p,q,r)a^*_{p+r}a^*_{-p}a_{q+r}a_{-q}.
	\end{aligned}
\]
Finally, $\cE_{M}$ contains terms where two operators have momenta in $P_H$ and two operators have momenta in $P_S.$ More precisely,
\[
	\begin{aligned}
		\cE_M=&\cE_{M,1}+\cE_{M,2}+\cE_{M,3}\\
		=&\frac1{N}\sum_{r\in \L^*}\sum_{\substack{p,q\in P_H:\\p+r,q+r\in P_S}}\varphi_1(p,q,r)a^*_pa^*_{q+r}a_qa_{p+r}\\
				&+\frac1N\sum_{r\in \L^*_+}\sum_{\substack{p,q\in P_H:\\p+r,q+r\in P_S}}\varphi_2(p,q,r)a^*_{p+r}a^*_{-p}a_{q+r}a_{-q}\\
				&+\frac1N\sum_{r\in \L^*}\sum_{\substack{p\in P_S,q\in P_H:\\p+r\in P_S,q+r\in P_H}}\varphi_3(p,q,r)a^*_pa^*_{q+r}a_qa_{p+r}
	\end{aligned}
\]
with
\[
		\begin{aligned}
			\varphi_1(p,q,r)=&N^\k\widehat V\Big(\frac r{N^{1-\k}}\Big)(\s_p\s_q\s_{p+r}\s_{q+r}+\g_p\g_q\g_{p+r}\g_{q+r}+2\g_p\g_{p+r}\s_q\s_{q+r})\\
			\varphi_2(p,q,r)=&N^\k\widehat{V}\Big(\frac r{N^{1-\k}}\Big)(\g_{p+r}\g_{q+r}\s_p\s_q\!+\!\g_p\g_{q+r}\s_{p+r}\s_q\!+\!\g_{p+r}\g_q\s_p\s_{q+r}\!+\!\g_p\g_q\s_{p+r}\s_{q+r})\\
			\varphi_3(p,q,r)=&N^\k\widehat{V}\Big(\frac r{N^{1-\k}}\Big)(\s_p\s_{p+r}\s_q\s_{q+r}\!+\!\g_p\g_q\g_{p+r}\g_{q+r}\!+\!\g_p\g_{p+r}\s_q\s_{q+r}\!+\!\s_q\s_{p+r}\g_p\g_{q+r}).
		\end{aligned}
\] 
Our goal is to prove
\be\label{eq:last}
	\begin{aligned}
		\frac{\langle\xi_\nu,\cE_2\xi_\nu\rangle}{\|\xi_\nu\|^2}\leq CN^{5\k/2-\e}
	\end{aligned}
\ee
for all $\k<2/3$ and $\e$ small enough.\\
Then, Eq. \eqref{eq:main} immediately follows by \eqref{eq:<xiGxi>}, \eqref{eq:en}, \eqref{eq:cEsplit}, \eqref{eq:errok} and \eqref{eq:last}.\\
To get the improved estimate \eqref{eq:last} the idea is to compute the expectation of each operator appearing on the right hand side of \eqref{eq:cE2} using the definition of $\xi_\nu.$ This is done in the rest of this section.

\medskip

{\it Remark} Note that by \eqref{eq:errok} and \eqref{eq:last} it follows that the quadratic conjugation only produces errors that remain small up to the thermodynamic limit, \ie for all $\k<2/3.$ Indeed, the restriction to $\k<7/12$ come from the action of the exponential of the cubic operator (see \eqref{eq:en}).

\medskip

\subsection{Bound of the expectation of \texorpdfstring{$\cE_C$}{EC} on \texorpdfstring{$\xi_\nu$}{xi}}
We start noting that the cubic error term $\cE_C$ has the same structure as the cubic term $\cC$ giving a large contribution to the energy analyzed in \cite[Sec. 5.2]{BCS}.\\
Recalling the definition of $A_\nu$ given in \eqref{eq:defA} 
we easily get
\[
	\begin{aligned}
		&\langle\xi_\nu,\cE_C\xi_\nu\rangle\\
		&=2\frac{\sqrt{N_0}}{N}\sum_{m\geq1}\frac1{m!(m-1)!}\sum_{\substack{p\in P_H,r\in P_S:\\p+r\in P_H}}\a(p,r)\langle A^m_\nu\O,a^*_{p+r}a^*_{-p}a^*_{-r}A^{m-1}_\nu\O\rangle\\
		&=2\sqrt{\frac{{N_0}}{N}}\sum_{m\geq 1}\frac{1}{m!(m-1)!N^m}\hspace{-0.2cm}\sum_{\substack{v_1,\tilde v_1\in P_S,r_1,\tilde r_1\in P_H:\\r_1+v_1,\tilde r_1+\tilde v_1\in P_H}}\hspace{-0.2cm}\cdots\hspace{-0.1cm}\sum_{\substack{v_m\in P_S,r_m\in P_H:\\r_m+v_m\in P_H}}\theta(\{r_j,v_j\}_{j=1}^m)\theta(\{\tilde r_j,\tilde v_j\}_{j=1}^{m-1})\\
		&\hphantom{{}={}}\times\!\prod_{i=1}^m\eta_{r_j}\s_{v_j}\!\!\prod_{j=1}^{m-1}\eta_{\tilde r_j}\s_{\tilde v_j}\hspace{-0.5cm}\sum_{\substack{p\in P_H,r\in P_S:\\p+r\in P_H}}\hspace{-0.5cm}\a(p,r)\langle \O,a_{r_m+v_m}\dots a_{-v_1}a^*_{p+r}a^*_{-p}a^*_{-r}a^*_{\tilde r_{m-1}+\tilde v_{m-1}}\dots a^*_{-\tilde v_1}\O\rangle.
	\end{aligned}
\]
Noting that the expectation in the last line vanishes unless there exists an index $i\in \{1,\dots,m\}$ such that $r=v_i$  and pairing the remaining  momenta in $P_S$ (by symmetry we can assume $r=-v_m$ and $\tilde v_j=v_j$ for all $j=1,\dots, m-1$) we find
\be
	\begin{aligned}
		&\langle\xi_\nu,\cE_C\xi_\nu\rangle\\
		&=2\sqrt{\frac{{N_0}}{N}}\sum_{m\geq 1}\frac{1}{(m-1)!N^m}\hspace{-0.2cm}\sum_{\substack{v_1\in P_S,r_1,\tilde r_1\in P_H:\\r_1+v_1,\tilde r_1+v_1\in P_H}}\hspace{-0.2cm}\cdots\hspace{-0.1cm}\sum_{\substack{v_m\in P_S,r_m\in P_H:\\r_m+v_m\in P_H}}\theta(\{r_j,v_j\}_{j=1}^m)\theta(\{\tilde r_j, v_j\}_{j=1}^{m-1})\\
		&\hphantom{{}={}}\times\prod_{j=1}^{m-1}\eta_{r_j}\eta_{\tilde r_j}\s_{v_j}^2\eta_{r_m}\s_{v_m}\sum_{\substack{p\in P_H:\\p+v_m\in P_H}}\a(p,v_m)\langle \O,A_{r_m,v_m}\dots A_{r_1,v_1}A^*_{p,v_m}\cdots A^*_{-\tilde r_1,v_1}\O\rangle\\
	\end{aligned}
\ee
where we introduced the notation $A_{r,v}=a_{r+v}a_{-r}$ for any $v\in P_S,r\in P_H$ such that $r+v\in P_H$ (and $A^*_{r,v}$ to denote the adjoint). Using now the fact that, due to the presence of $\theta(\{r_j,v_j\}_{j=1}^m)\theta(\{\tilde r_j,v_j\}_{j=1}^{m-1}),$ each $A_{r_j,v_j}$ has to be contracted with $A^*_{\tilde r_j,v_j}$ for any $j=1,\dots, m-1$ and therefore $A_{r_m,v_m}$ is contracted with $A^*_{p,v_m}$ we get
\be\label{eq:cECbound}
\begin{aligned}
		&\langle\xi_\nu,\cE_C\xi_\nu\rangle\\
		&=2\sqrt{\frac{{N_0}}{N}}\sum_{m\geq 1}\frac{1}{(m-1)!N^m}\hspace{-0.2cm}\sum_{\substack{v_1\in P_S,r_1,\tilde r_1\in P_H:\\r_1+v_1,\tilde r_1+v_1\in P_H}}\hspace{-0.2cm}\cdots\hspace{-0.1cm}\sum_{\substack{v_m\in P_S,r_m\in P_H:\\r_m+v_m\in P_H}}\theta(\{r_j,v_j\}_{j=1}^m)\theta(\{\tilde r_j, v_j\}_{j=1}^{m-1})\\
		&\hphantom{{}={}}\times\prod_{j=1}^{m-1}\eta_{r_j}\eta_{\tilde r_j}(\d_{\tilde r_j,r_j}+\d_{-\tilde r_j,r_j+v_j})\s_{v_j}^2\eta_{r_m}\s_{v_m}\sum_{\substack{p\in P_H:\\p+v_m\in P_H}}\a(p,v_m)\sum_{p_m\in \{-r_m,r_m+v_m\}}\d_{-p,p_m}.
	\end{aligned}
\ee
We now note that 
\[
	|\a(p,v_m)\d_{p,-p_m}|\leq CN^\k(\g_{v_m}|\s_{-p_m+v_m}|+|\s_{v_m}||\s_{p_m}|).
\]
Takng the absolute value in \eqref{eq:cECbound} and using the fact that when all terms in the sum are positive we can replace $\theta(\{r_j,v_j\}_{j=1}^{m})$ with $\theta(\{r_j,v_j\}_{j=1}^{m-1})$ obtaining an upper bound we conclude by \eqref{eq:norm_xi}, \eqref{eq:eta_H} and \eqref{eq:tau_S}
\be\label{eq:cEC}
	\begin{aligned}
		\frac{|\langle\xi_\nu,\cE_C\xi_\nu\rangle|}{\|\xi_\nu\|^2}
		\leq&CN^{\k-1}\|\eta_H\|^2(\|\g_S\s_S\|_1+\|\s_S\|^2)\leq C N^{5\k/2-\e}
	\end{aligned}
\ee
for all $\k<2/3$ and $\e$ small enough.
\subsection{Bound of the expectation of \texorpdfstring{$\cE_{H}$}{EH} on \texorpdfstring{$\xi_\nu$}{xi}}
To bound the quartic error term $\cE_H,$ we start considering $\cE_{H,1}$ which has the same form as the large quartic term $\cV^{(H)}_N$ (see \cite[Sec. 5.3]{BCS}).\\
We write,
\[
	\begin{aligned}
		&\langle\xi_\nu,\cE_{H,1}\xi_\nu\rangle\\
		&=\sum_{m\geq1}\frac1{m!}\frac1{2N^{m+1}}\hspace{-0.2cm}\sum_{\substack{v_1\in P_S,r_1,\tilde{r}_1\in P_H:\\r_1+v_1,\tilde{r}_1+v_1\in P_H}}\hspace{-0.2cm}\cdots\hspace{-0.1cm}\sum_{\substack{v_m\in P_S,r_m,\tilde{r}_m\in P_H:\\r_m+v_m,\tilde{r}_m+v_m\in P_H}}\hspace{-0.3cm}\theta(\{r_j,v_j\}_{j=1}^m)\theta(\{\tilde{r}_j,v_j\}_{j=1}^m)\prod_{j=1}^m\eta_{r_j}\eta_{\tilde{r}_j}\s_{v_j}^2\\
		&\hphantom{{}={}}\times\sum_{r\in \L^*}\sum_{\substack{p,q\in P_H:\\p+r,q+r\in P_H}}\b_1(p,q,r)\langle\O,A_{r_1,v_1}\cdots A_{r_m,v_m}a^*_{p}a^*_{q+r}a_{q}a_{p+r}A^*_{\tilde{r}_1,v_1}\cdots A^*_{\tilde{r}_m,v_m}\O\rangle
	\end{aligned}
\]
where we paired all momenta in $P_S$. 
\\We now distinguish two contributions: the first one corresponds to the situation in which $a_q,a_{p+r}$ are annihilated with $A^*_{\tilde r_i,v_i}$ for some $i=1,\dots,m$ (this also implies, taking into account the presence of $\theta(\{r_j,v_j\}_{j=1}^m)\theta(\{\tilde{r}_j,v_j\}_{j=1}^m)$, that $a^*_p,a^*_{q+r}$ are annihilated with $A_{r_i,v_i}$). The second case on the other hand,  arises when $a_q,a_{p+r}$ are annihilated with $a^*_{\tilde p_i}, a^*_{\tilde p_j}$ for $\tilde p_\ell\in \{-\tilde r_\ell, \tilde r_\ell+v_\ell\},$ $\ell=i,j$ with $i\neq j$ (then $a^*_p,a^*_{q+r}$ are annihilated with $a_{p_i}, a_{p_j},$ for $p_\ell\in \{-r_\ell,r_\ell+v_\ell\},$ $\ell=i,j$ and $a^*_{-\tilde p_i+v_i},a^*_{-\tilde p_j+v_j}$ are annihilated with $a_{-p_i+v_i}, a_{-p_j+v_j}$).
\\We denote the two contributions just described by $A$ and $B$ respectively so that
\be\label{eq:cEH1split}
	\langle\xi_\nu,\cE_{H,1}\xi_\nu\rangle=A+B
\ee
with
\[
	\begin{aligned}
		A=&\sum_{m\geq 1}\frac1{(m-1)!}\frac1{2N^{m+1}}\!\!\!\!\sum_{\substack{v_1 \in P_S, r_1, \tl r_1 \in P_H: \\ r_1 + v_1 , \tl r_1 + v_1 \in P_H}}\!\! \cdots\!\! \sum_{\substack{v_{m} \in P_S, r_{m}, \tl r_{m} \in P_H: \\ r_{m} + v_{m} , \tl r_{m} + v_{m} \in P_H}}\hskip -0.7cm\theta\big( \{r_j, v_j \}_{j=1}^{m} \big) \theta\big( \{\tl r_j, v_j \}_{j=1}^{m} \big)\\[6pt]
&\, \times \hskip -0.1cm \prod_{j=1}^{m}\eta_{r_j}\eta_{\tl r_j}\s_{v_j}^2\!\!\prod_{j=1}^{m-1}\hskip -0.1cm(\d_{r_j,\tl r_j}+\d_{-r_j,\tl r_j+v_j})\sum_{r\in \L^*}\sum_{p\in P_H}\! \b_1(p,v_m-p-r,r) \hskip -0.7cm \sum_{\substack{p_m\in \{-r_m,r_m+v_m\}\\\tl p_m\in \{-\tl r_m,\tl r_m+v_m\}}}\hspace{-0.8cm}\d_{p,p_m}\d_{p+r,\tl p_m} \\
	B=& \sum_{m\geq 2}\frac1{(m-2)!}\frac1{2N^{m+1}} \sum_{\substack{v_1 \in P_S\,, r_1, \tl r_1 \in P_H: \\ r_1 + v_1 ,\, \tl r_1 +  v_1 \in P_H}} \cdots 
\sum_{\substack{v_{m} \in P_S\,, r_{m}, \tl r_{m} \in P_H: \\ r_{m} + v_{m} ,\, \tl r_{m} +  v_{m} \in P_H}}\hskip -0.5cm\theta\big( \{r_j, v_j \}_{j=1}^{m} \big) \theta\big( \{\tl r_j, v_j \}_{j=1}^{m} \big)  \\
& \times \prod_{i=1}^{m}\eta_{r_i}\eta_{\tilde{r}_i}\s_{v_i}^2\prod_{j=1}^{m-2} \big( \d_{\tl r_i,r_i}+\d_{-\tl r_i,r_i+v_i}\big)  \sum_{r \in \L^*}\sum_{p,q \in P_H} \b_1(p,q,r)\!\!\!   \sum_{\substack{ p_\ell \in \{ - r_\ell, r_\ell+v_\ell\}  \\ \ell=m-1,m } }\!\!\!\d_{p,  p_m} \d_{q+r,  p_{m-1}} \\
& \hspace{0.5cm}  \times  \hskip -0.5cm 
\sum_{\substack{ \tl p_\ell \in \{ -\tl r_\ell,  \tl r_\ell+v_\ell\}   \\ \ell=m-1,m } }  \big(\d_{q, \tl p_m} \d_{p+r,  \tl p_{m-1}} +  \d_{q, \tl p_{m-1}} \d_{p+r,\tl p_{m}} \big)  \big( \d_{\tl p_m, p_m} + \d_{-\tl p_m+v_m, -p_{m-1}+v_{m-1}} \big) .
	\end{aligned}
\]
Thus, the bound
\[
	\begin{aligned}
		 |\b_1(p,v_m-p-r,r)\d_{p,p_m}\d_{p+r,\tl p_m}|\leq& CN^\k\widehat{V}\Big(\frac{r}{N^{1-\k}}\Big)(|\eta_p|^3+|\eta_{p+r}|^3+|\eta_{p-v_m}|^3+|\eta_{p+r-v_m}|^3)\\
		 &+CN^\k(\|\eta_H\|_\infty^2|\eta_p||\eta_{p+r}|+|\eta_{p-v_m}||\eta_{p+r-v_m}|)
	\end{aligned}
\]
yields
\be\label{eq:A}
	\begin{aligned}
		\frac{A}{\|\xi_\nu\|^2}\leq& CN^{\k-2}\|\s_S\|^2(N\|\eta_H\|_\infty^2\|\eta_H\|^2+\|\eta_H\|_\infty^2\|\eta_H\|^4+\|\eta_H\|^4)\\
	\end{aligned}
\ee
where we used the bound $\sup_{r\in \L^*}\sum_{p\in P_H}N^\k\widehat V(r/N^{1-\k})/|p-r|^{2}\leq N$ and the fact that $|\eta_p|\leq CN^\k|p|^{-2}$.\\
On the other hand, since $\sup_{\substack{r\in \L^*,p,q\in P_H:\\p+r,q+r\in P_H}}|\b_1(p,q,r)|\leq CN^\k\|\eta_H\|_\infty^2,$ we get
\be\label{eq:B}
	\begin{aligned}
		\frac{|B|}{\|\xi_\nu\|^2}
			\leq &CN^{\k-3}\|\s_S\|^4\|\eta_H\|^4\|\eta_H\|_\infty^2.
	\end{aligned}
\ee
We now consider $\cE_{H,2}.$ First, pairing all momenta in $P_S$, we rewrite
\[
	\begin{aligned}
		&\langle\xi_\nu,\cE_{H,2}\xi_\nu\rangle\\
		&=\sum_{m\geq1}\frac1{m!}\frac1{N^{m+1}}\hspace{-0.2cm}\sum_{\substack{v_1\in P_S,r_1,\tilde{r}_1\in P_H:\\r_1+v_1,\tilde{r}_1+v_1\in P_H}}\hspace{-0.2cm}\cdots\hspace{-0.1cm}\sum_{\substack{v_m\in P_S,r_m,\tilde{r}_m\in P_H:\\r_m+v_m,\tilde{r}_m+v_m\in P_H}}\hspace{-0.3cm}\theta(\{r_j,v_j\}_{j=1}^m)\theta(\{\tilde{r}_j,v_j\}_{j=1}^m)\prod_{j=1}^m\eta_{r_j}\eta_{\tilde{r}_j}\s_{v_j}^2\\
		&\hphantom{{}={}}\times\sum_{r\in \L^*}\sum_{\substack{p,q\in P_H:\\p+r,q+r\in P_H}}\b_2(p,q,r)\langle\O,A_{r_1,v_1}\cdots A_{r_m,v_m}a^*_{p+r}a^*_{-p}a_{q+r}a_{-q}A^*_{\tilde{r}_1,v_1}\cdots A^*_{\tilde{r}_m,v_m}\O\rangle.
	\end{aligned}
\]
We note that also in this case we can distinguish two contributions, depending on whether the operators $a_{q+r}a_{-q}$ are annihilated with $A^*_{\tilde{r}_j,v_j}$ or with $a^*_{\tilde{p}_j},a^*_{\tilde{p}_k},$ with $\tl p_\ell\in\{-\tl r_\ell ,\tl r_\ell+v_\ell\},$ $\ell=j,k$ and $j\neq k.$ \\
Hence, we split
\be\label{eq:cEH2split}
	\langle\xi_\nu,\cE_{H,2}\xi_\nu\rangle=C+D
\ee
with $C,D$ defined by
\[
	\begin{aligned}
		C=&\sum_{m\geq1}\frac1{(m-1)!}\frac1{N^{m+1}}\hspace{-0.3cm}\sum_{\substack{v_1\in P_S,r_1,\tilde{r}_1\in P_H:\\r_1+v_1,\tilde{r}_1+v_1\in P_H}}\hspace{-0.2cm}\cdots\hspace{-0.1cm}\sum_{\substack{v_m\in P_S,r_m,\tilde{r}_m\in P_H:\\r_m+v_m,\tilde{r}_m+v_m\in P_H}}\hspace{-0.6cm}\theta(\{r_j,v_j\}_{j=1}^m)\theta(\{\tilde{r}_j,v_j\}_{j=1}^m)\eta_{r_m}\eta_{\tilde{r}_m}\s_{v_m}^2\\
		&\times \prod_{j=1}^{m-1}\eta_{r_j}\eta_{\tilde{r}_j}\s_{v_j}^2(\d_{\tilde{r}_j,r_j}+\d_{-\tilde{r}_j,r_j+v_j})\hspace{-0.2cm} \sum_{p,q\in P_H}\b_2(p,q,v_m)\hspace{-0.2cm}\sum_{\substack{p_m\in \{-r_m,r_m+v_m\}\\\tilde p_m\in \{-\tilde{r}_m,\tilde{r}_m+v_m\}}}\hspace{-0.2cm}\d_{-p,p_m}\d_{-q,\tilde{p}_m}\\
		D=&\sum_{m\geq 2}\frac1{(m-2)!}\frac1{N^{m+1}}\sum_{\substack{v_1\in P_S,r_1,\tilde{r}_1\in P_H:\\r_1+v_1,\tilde{r}_1+v_1\in P_H}}\cdots\sum_{\substack{v_m\in P_S,r_m,\tilde{r}_m\in P_H:\\r_m+v_m,\tilde{r}_m+v_m\in P_H}}\theta(\{r_j,v_j\}_{j=1}^m)\theta(\{\tilde{r}_j,v_j\}_{j=1}^m)\\
		&\times \prod_{j=1}^{m}\eta_{r_j}\eta_{\tilde{r}_j}\s_{v_j}^2\prod_{i=1}^{m-2}(\d_{\tilde{r}_j,r_j}+\d_{-\tilde{r}_j,r_j+v_j}) \sum_{r\in \L^*}\sum_{p,q\in P_H}\b_2(p,q,r)\!\!\sum_{\substack{p_\ell\in\{-r_\ell,r_\ell+v_\ell\}\\\tilde{p}_\ell\in\{-\tilde{r}_\ell,\tilde{r}_\ell+v_\ell\}\\\ell=m-1,m}}\hspace{-0.5cm}\d_{-q,\tilde{p}_m}\d_{q+r,\tilde{p}_{m-1}}\\
		&\hspace{2cm}\times(\d_{-p,p_m}\d_{p+r,p_{m-1}}+\d_{-p,p_{m-1}}\d_{p+r,p_m})(\d_{p_m,\tilde{p}_m}+\d_{-\tilde{p}_m+v_m,-p_{m-1}+v_{m-1}}).
	\end{aligned}
\]
Taking the absolute value, which allows us to forget $\theta(\{\tilde r_j,v_j\}_{j=1}^m)$ and replace $\theta(\{r_j,v_j\}_{j=1}^m)$ with $\theta(\{r_j,v_j\}_{j=1}^{m-1}),$ and noting that $|\b_2(p,q,r)|\leq CN^\k|\eta_p||\eta_q|$ for any $r\in \L^*,p,q\in P_H$ with $p+r,q+r\in P_H$ we obtain
\be\label{eq:C}
	\frac{|C|}{\|\xi_\nu\|^2}\leq CN^{\k-2}\|\s_S\|^2\|\eta_H\|^4
\ee
and
\be\label{eq:D}
	\frac{|D|}{\|\xi_\nu\|^2}\leq CN^{\k-3}\|\s_S\|^4\|\eta_H\|^4\|\eta_H\|_\infty^2.
\ee
By \eqref{eq:cEH1split} and \eqref{eq:cEH2split}, using the bounds \eqref{eq:A},\eqref{eq:B},\eqref{eq:C},\eqref{eq:D} and recalling \eqref{eq:eta_H} and \eqref{eq:tau_S} we get
\be\label{eq:cEH}
	\frac{|\langle\xi_\nu,\cE_H\xi_\nu\rangle|}{\|\xi_\nu\|^2}\leq CN^{5\k/2-\e}
\ee
for any $\k<2/3$ and $\e$ small enough.
\subsection{Bound of the expectation of \texorpdfstring{$\cE_{S}$}{ES} on \texorpdfstring{$\xi_\nu$}{xi}}
In this subsection we focus on $\cE_S.$ Let us first consider $\cE_{S,1};$ by definition
\be\label{eq:E_S1}
	\begin{aligned}
		\langle\xi_\nu,\cE_{S,1}\xi_\nu\rangle=&\sum_{m\geq 2}\frac1{(m!)^2}\frac{1}{2N^{m+1}}\sum_{\substack{v_1,\tilde{v}_1\in P_S,r_1,\tilde{r}_1\in P_H:\\r_1+v_1,\tilde{r}_1+\tilde{v}_1\in P_H}}\cdots\sum_{\substack{v_m,\tilde{v}_m\in P_S,r_m,\tilde{r}_m\in P_H:\\r_m+v_m,\tilde{r}_m+\tilde{v}_m\in P_H}}\\
		&\times\theta\big(\{r_j,v_j\}_{j=1}^m\big)\theta\big(\{\tilde{r}_j,\tilde{v}_j\}_{j=1}^m\big)\prod_{j=1}^m\eta_{r_j}\eta_{\tilde{r}_j}\s_{v_j}\s_{\tilde{v}_j}\sum_{r\in\L^*}\sum_{\substack{p,q\in P_S:\\p+r,q+r\in P_S}}\zeta_1(p,q,r)\\
		&\times\langle \O, a_{r_m+v_m}a_{-r_m}a_{-v_m}\cdots a_{-v_1}a^*_{p+r}a^*_qa_pa_{q+r}a^*_{\tilde{r}_m+\tilde{v}_m}a^*_{-\tilde{r}_m}a^*_{-\tilde{v}_m} \cdots a^*_{-\tilde{v}_1}\O\rangle
	\end{aligned}
\ee
 We note that the scalar product in the last line of \eqref{eq:E_S1} does not vanish only if there exist $i,j,k,\ell$ such that $q=-\tilde{v}_i,p+r=-\tilde{v}_j,p=-v_k, q+r=-v_\ell$ which immediately implies $r=v_k-\tilde{v}_j$ and $\tilde{v}_i=v_k+v_\ell-\tilde{v}_j.$ By symmetry we can assume $i=k=m$ and $j=\ell=m-1$ getting a factor $m^2(m-1)^2$ in front. Pairing also the remaining $m-2$ momenta in $P_S$ we obtain
\[
	\begin{aligned}
		\langle\xi_\nu,&\cE_{S,1}\xi_\nu\rangle\\
		=&\sum_{m\geq 2}\frac1{(m-2)!}\frac{1}{2N^{m+1}}\!\!\!\sum_{\substack{v_1,\tilde{v}_1\in P_S,r_1,\tilde{r}_1\in P_H:\\r_1+v_1,\tilde{r}_1+\tilde{v}_1\in P_H}}\!\!\!\cdots\!\!\sum_{\substack{v_m,\tilde{v}_m\in P_S,r_m,\tilde{r}_m\in P_H:\\r_m+v_m,\tilde{r}_m+\tilde{v}_m\in P_H}}\zeta_1(-v_m,-\tilde{v}_m,v_m-\tilde{v}_{m-1})\\
		&\times\theta\big(\{r_j,v_j\}_{j=1}^m\big)\theta\big(\{\tilde{r}_j,\tilde{v}_j\}_{j=1}^m\big)\prod_{j=1}^m\eta_{r_j}\eta_{\tilde{r}_j}\s_{v_j}\s_{\tilde{v}_j}\d_{\tilde{v}_m,v_m+{v}_{m-1}-\tilde v_{m-1}}\prod_{i=1}^{m-2}\d_{\tilde v_i,v_i}\\
		&\times\langle \O,A_{r_m,v_m}\cdots A_{r_1,v_1}A^*_{\tilde r_m,v_m+v_{m-1}-\tilde{v}_{m-1}} A^*_{\tilde{r}_{m-1},\tilde{v}_{m-1}}A^*_{\tilde r_{m-2},v_{m-2}}\cdots A^*_{\tilde r_1,v_1}\O\rangle.
	\end{aligned}
\]
We now distinguish three contributions. The first contribution, which we will denote by $I$,  corresponds to the situation in which the operators in $A_{r_m,v_m}$ and $A_{r_{m-1},v_{m-1}}$ are annihilated with the operators in $A^*_{\tilde{r}_m,v_m+v_{m-1}-\tilde v_{m-1}}$ and in $A^*_{\tilde{r}_{m-1},\tilde v_{m-1}}$ respectively (note that this immediately implies $v_{m-1}=\tilde{v}_{m-1}$). The second contribution, denoted by $II$, is on the contrary obtained when $A_{r_m,v_m}$ is annihilated with $A^*_{\tilde{r}_{m-1},\tilde v_{m-1}}$ and $A_{r_{m-1},v_{m-1}}$ is annihilated with $A^*_{\tilde{r}_m,v_m+v_{m-1}-\tilde v_{m-1}}$ (then $\tilde{v}_{m-1}=v_m$). Finally, there is a third term, denoted by $III$, arising when one operator in $A_{r_m,v_m}$ is annihilated with an operator in $A^*_{\tilde{r}_{m-1},\tilde{v}_{m-1}}$ and the other with an operator in $A^*_{\tilde{r}_m,v_m+v_{m-1}-\tilde{v}_{m-1}}$ (and analogously an operator in $A_{r_{m-1},v_{m-1}}$ is annihilated with an operaotr in $A^*_{\tilde{r}_{m-1},\tilde{v}_{m-1}}$ and the other with an operator in $A^*_{\tilde{r}_m,v_m+v_{m-1}-\tilde{v}_{m-1}}$). Let us stress that in all these cases the operators $A_{r_j,v_j}$ are annihilated with the operators $A^*_{\tilde{r}_j,v_j}$ for any $j=1,\dots,m-2$ due to the presence of the restrictions encoded in $\theta(\{r_j,v_j\}_{j=1}^m)\theta(\{\tilde{r}_j,\tilde{v}_j\}_{j=1}^m).$ \\
Summarizing, we have $\langle\xi_\nu,\cE_{S,1}\xi_\nu\rangle=I+II+III$ with 
\[
	\begin{aligned}
		I=&\sum_{m\geq 2}\frac1{(m-2)!}\frac{1}{2N^{m+1}}\hspace{-0.2cm}\sum_{\substack{v_1\in P_S,r_1,\tilde{r}_1\in P_H:\\r_1+v_1,\tilde{r}_1+{v}_1\in P_H}}\hspace{-0.2cm}\cdots\hspace{-0.1cm}\sum_{\substack{v_m\in P_S,r_m,\tilde{r}_m\in P_H:\\r_m+v_m,\tilde{r}_m+{v}_m\in P_H}}\hspace{-0.3cm}\zeta_1(-v_m,-{v}_m,v_m-{v}_{m-1})\\
		&\times\theta\big(\{r_j,v_j\}_{j=1}^m\big)\theta\big(\{\tilde{r}_j,{v}_j\}_{j=1}^m\big)\prod_{j=1}^m\eta_{r_j}\eta_{\tl r_j}\s_{v_j}^2(\d_{\tl r_j,r_j}+\d_{-\tl r_j,r_j+v_j})\\
		II=&\sum_{m\geq 2}\frac1{(m-2)!}\frac{1}{2N^{m+1}}\hspace{-0.2cm}\sum_{\substack{v_1,\tilde{v}_1\in P_S,r_1,\tilde{r}_1\in P_H:\\r_1+v_1,\tilde{r}_1+\tilde{v}_1\in P_H}}\hspace{-0.2cm}\cdots\hspace{-0.1cm}\sum_{\substack{v_m,\tilde{v}_m\in P_S,r_m,\tilde{r}_m\in P_H:\\r_m+v_m,\tilde{r}_m+\tilde{v}_m\in P_H}}\hspace{-0.2cm}\zeta_1 (-v_m,-\tilde{v}_m,v_m-\tilde{v}_{m-1})\\
		&\times\theta\big(\{r_j,v_j\}_{j=1}^m\big)\theta\big(\{\tilde{r}_j,\tilde{v}_j\}_{j=1}^m\big)\prod_{j=1}^m\eta_{r_j}\eta_{\tilde{r}_j}\s_{v_j}\s_{\tilde{v}_j}\d_{\tilde{v}_m,v_m+{v}_{m-1}-\tilde v_{m-1}}\\
		&\times\prod_{i=1}^{m-2}\d_{\tilde v_i,v_i}(\d_{\tilde r_i,r_i}+\d_{-\tilde{r}_i,r_i+v_i})\d_{\tilde{v}_{m-1},v_{m}}\sum_{\substack{p_\ell\in \{-r_\ell,r_\ell+v_\ell\}\\\tilde{p}_\ell\in \{-\tilde{r}_\ell,\tilde{r}_\ell+\tilde{v}_\ell\}\\ \ell=m-1,m}}\d_{\tilde{p}_{m-1},p_{m}}\d_{\tilde{p}_m,p_{m-1}}\\
		III=&\sum_{m\geq 2}\frac1{(m-2)!}\frac{1}{2N^{m+1}}\hspace{-0.2cm}\sum_{\substack{v_1,\tilde{v}_1\in P_S,r_1,\tilde{r}_1\in P_H:\\r_1+v_1,\tilde{r}_1+\tilde{v}_1\in P_H}}\hspace{-0.2cm}\cdots\hspace{-0.1cm}\sum_{\substack{v_m,\tilde{v}_m\in P_S,r_m,\tilde{r}_m\in P_H:\\r_m+v_m,\tilde{r}_m+\tilde{v}_m\in P_H}}\hspace{-0.1cm}\zeta_1(-v_m,-\tilde{v}_m,v_m-\tilde{v}_{m-1})\\
		&\times\theta\big(\{r_j,v_j\}_{j=1}^m\big)\theta\big(\{\tilde{r}_j,\tilde{v}_j\}_{j=1}^m\big)\prod_{j=1}^m\eta_{r_j}\eta_{\tilde{r}_j}\s_{v_j}\s_{\tilde{v}_j}\d_{\tilde{v}_m,v_m+{v}_{m-1}-\tilde v_{m-1}}\\
		&\times\prod_{i=1}^{m-2}\d_{\tilde v_i,v_i}(\d_{\tilde r_i,r_i}+\d_{-\tilde{r}_i,r_i+v_i})\sum_{\substack{p_\ell\in \{-r_\ell,r_\ell+v_\ell\}\\\tilde{p}_\ell\in \{-\tilde{r}_\ell,\tilde{r}_\ell+\tilde{v}_\ell\}\\ \ell=m-1,m}}\d_{\tilde{p}_{m-1},p_{m-1}}\d_{\tilde{p}_m,p_m}\d_{-\tilde{p}_{m-1}+\tilde{v}_{m-1},-p_m+v_m}.\\
	\end{aligned}
\]
Recalling the definition of $\zeta_1$ and using \eqref{eq:tau_S} we find $|\zeta_1(p,q,r)|\leq CN^{\k+2\e}$ for any $r\in \L^*,p,q\in P_S$ such that $p+r,q+r\in P_S$. Thus,
\be\label{eq:I}
	\begin{aligned}
		\frac{|I|+|II|}{\|\xi_\nu\|^2}\leq
		CN^{\k-3+2\e}\|\s_S\|^4\|\eta_H\|^4.
	\end{aligned}
\ee
To bound $III$ we note that
\[
	\begin{aligned}
	&|\d_{\tilde{v}_m,v_m+{v}_{m-1}-\tilde v_{m-1}}\zeta_1(-v_m,-\tilde{v}_m,v_m-\tilde{v}_{m-1})|\leq CN^{\k}\\
	&\times(|\s_{v_m}||\s_{\tilde v_{m-1}}||\s_{v_{m-1}}|\|\s_S\|_\infty+|\g_{v_m}||\g_{\tilde v_{m-1}}||\g_{v_{m-1}}|\|\g_S\|_\infty+|\g_{v_m}||\g_{\tilde v_{m-1}}||\s_{v_{m-1}}|\|\s_S\|_\infty).
	\end{aligned}
\]
Hence,
\be\label{eq:III}
	\begin{aligned}
		\frac{| III |}{\|\xi_\nu\|^2}&\leq CN^{\k-3}\|\eta_H\|^2\|\eta_H\|_\infty^2\|\s_S\|_\infty\\
		&\hspace{0.4cm}\times(\|\s_S\|^6\|\s_S\|_\infty+\|\g_S\s_S\|_1^3\|\g_S\|_\infty+\|\s_S\|^2\|\g_S\s_S\|_1^2\|\s_S\|_\infty)\\
	\end{aligned}
\ee
From \eqref{eq:I} and \eqref{eq:III}, unsing \eqref{eq:eta_H},\eqref{eq:tau_S} we conclude
\be\label{eq:S1}
	\frac{|\langle\xi_\nu,\cE_{S,1}\xi_\nu\rangle|}{\|\xi_\nu\|^2}\leq\frac{|I|+|II|+|III|}{\|\xi_\nu\|^2}\leq CN^{5\k/2-\e}
\ee
for any $\k<2/3$ and $\e$ small enough.\\
The error term $\cE_{S,2}$ can be bounded analogously. Indeed, reasoning as before we split
\[
	\langle\xi_\nu,\cE_{S,2}\xi_\nu\rangle=\widetilde I+\widetilde{II}+\widetilde{III}
\]
where
\[
	\begin{aligned}
		\widetilde I=&\sum_{m\geq 2}\frac1{(m-2)!}\frac{1}{N^{m+1}}\!\!\!\sum_{\substack{v_1\in P_S,r_1,\tilde{r}_1\in P_H:\\r_1+v_1,\tilde{r}_1+{v}_1\in P_H}}\!\!\!\cdots\!\!\sum_{\substack{v_m\in P_S,r_m,\tilde{r}_m\in P_H:\\r_m+v_m,\tilde{r}_m+{v}_m\in P_H}}\zeta_2(v_m,{v}_m,-v_m-v_{m-1})\\
		&\times\theta\big(\{r_j,v_j\}_{j=1}^m\big)\theta\big(\{\tilde{r}_j,\tilde{v}_j\}_{j=1}^m\big)\prod_{j=1}^m\eta_{r_j}\eta_{\tl r_j}\s_{v_j}^2(\d_{\tl r_j,r_j}+\d_{-\tl r_j,r_j+v_j})\\
		\widetilde{II}=&\sum_{m\geq 2}\frac1{(m-2)!}\frac{1}{N^{m+1}}\!\!\!\sum_{\substack{v_1,\tilde{v}_1\in P_S,r_1,\tilde{r}_1\in P_H:\\r_1+v_1,\tilde{r}_1+\tilde{v}_1\in P_H}}\!\!\!\cdots\!\!\sum_{\substack{v_m,\tilde{v}_m\in P_S,r_m,\tilde{r}_m\in P_H:\\r_m+v_m,\tilde{r}_m+\tilde{v}_m\in P_H}}\zeta_2 (v_m,\tilde{v}_m,-v_m-v_{m-1})\\
		&\times\theta\big(\{r_j,v_j\}_{j=1}^m\big)\theta\big(\{\tilde{r}_j,\tilde{v}_j\}_{j=1}^m\big)\prod_{j=1}^m\eta_{r_j}\eta_{\tilde{r}_j}\s_{v_j}\s_{\tilde{v}_j}\d_{\tilde{v}_m,v_m+{v}_{m-1}-\tilde v_{m-1}}\\
		&\times\prod_{i=1}^{m-2}\d_{\tilde v_i,v_i}(\d_{\tilde r_i,r_i}+\d_{-\tilde{r}_i,r_i+v_i})\d_{\tilde{v}_{m-1},v_{m}}\sum_{\substack{p_\ell\in \{-r_\ell,r_\ell+v_\ell\}\\\tilde{p}_\ell\in \{-\tilde{r}_\ell,\tilde{r}_\ell+\tilde{v}_\ell\}\\ \ell=m-1,m}}\d_{\tilde{p}_{m-1},p_{m}}\d_{\tilde{p}_m,p_{m-1}}\\
		\widetilde{III}=&\sum_{m\geq 2}\frac1{(m-2)!}\frac{1}{N^{m+1}}\!\!\!\sum_{\substack{v_1,\tilde{v}_1\in P_S,r_1,\tilde{r}_1\in P_H:\\r_1+v_1,\tilde{r}_1+\tilde{v}_1\in P_H}}\!\!\!\cdots\!\!\sum_{\substack{v_m,\tilde{v}_m\in P_S,r_m,\tilde{r}_m\in P_H:\\r_m+v_m,\tilde{r}_m+\tilde{v}_m\in P_H}}\zeta_2(v_m,\tilde{v}_m,-v_m-v_{m-1})\\
		&\times\theta\big(\{r_j,v_j\}_{j=1}^m\big)\theta\big(\{\tilde{r}_j,\tilde{v}_j\}_{j=1}^m\big)\prod_{j=1}^m\eta_{r_j}\eta_{\tilde{r}_j}\s_{v_j}\s_{\tilde{v}_j}\d_{\tilde{v}_m,v_m+{v}_{m-1}-\tilde v_{m-1}}\\
		&\times\prod_{i=1}^{m-2}\d_{\tilde v_i,v_i}(\d_{\tilde r_i,r_i}+\d_{-\tilde{r}_i,r_i+v_i})\sum_{\substack{p_\ell\in \{-r_\ell,r_\ell+v_\ell\}\\\tilde{p}_\ell\in \{-\tilde{r}_\ell,\tilde{r}_\ell+\tilde{v}_\ell\}\\ \ell=m-1,m}}\d_{\tilde{p}_{m-1},p_{m-1}}\d_{\tilde{p}_m,p_m}\d_{-\tilde{p}_{m-1}+\tilde{v}_{m-1},-p_m+v_m}.
	\end{aligned}
\]
Hence, noting that $|\zeta_2(p,q,r)|\leq CN^\k|\g_{p+r}|\g_{q+r}||\s_p||\s_q|$, we find
\be\label{eq:Itilde}
	\begin{aligned}
		\frac{|\widetilde I|+|\widetilde{II}|}{\|\xi_\nu\|^2}\leq
		CN^{\k-3+2\e}\|\s_S\|^4\|\eta_H\|^4
	\end{aligned}
\ee
and
\be\label{eq:IIItilde}
		\frac{| \widetilde{III}|}{\|\xi_\nu\|^2}\leq CN^{\k-3}\|\eta_H\|^2\|\eta_H\|_\infty^2\|\s_S\|_\infty^2\|\g_S\s_S\|_1^2\|\s_S\|^2
\ee
Using \eqref{eq:eta_H} and \eqref{eq:tau_S} we get from \eqref{eq:S1}, \eqref{eq:Itilde} and \eqref{eq:IIItilde} that
\be\label{eq:cES}
	\frac{|\langle\xi_\nu,\cE_{S}\xi_\nu\rangle|}{\|\xi_\nu\|^2}\leq CN^{5\k/2-\e}
\ee
for any $\k<2/3$ and $\e$ small enough.
\subsection{Bound of the expectation of \texorpdfstring{$\cE_M$}{EM} on \texorpdfstring{$\xi_\nu$}{xi}}
To conclude the proof of \eqref{eq:last} it remains to study $\cE_{M}.$ We start focusing on $\cE_{M,1};$ we
rewrite
\[
	\begin{aligned}
		\langle\xi_\nu,\cE_{M,1}\xi_\nu\rangle=&\sum_{m\geq 1}\frac1{(m!)^2}\frac{1}{N^{m+1}}\sum_{\substack{v_1,\tilde{v}_1\in P_S,r_1,\tilde{r}_1\in P_H:\\r_1+v_1,\tilde{r}_1+\tilde{v}_1\in P_H}}\cdots\sum_{\substack{v_m,\tilde{v}_m\in P_S,r_m,\tilde{r}_m\in P_H:\\r_m+v_m,\tilde{r}_m+\tilde{v}_m\in P_H}}\\
		&\times\theta\big(\{r_j,v_j\}_{j=1}^m\big)\theta\big(\{\tilde{r}_j,\tilde{v}_j\}_{j=1}^m\big)\prod_{j=1}^m\eta_{r_j}\eta_{\tilde{r}_j}\s_{v_j}\s_{\tilde{v}_j}\sum_{r\in \L^*}\sum_{\substack{p,q\in P_H:\\p+r,q+r\in P_S}}\varphi_1(p,q,r)\\
		&\times\langle \O, a_{r_m+v_m}\cdots a_{-v_1}a^*_{p+r}a^*_qa_pa_{q+r}a^*_{\tilde{r}_m+\tilde{v}_m} \cdots a^*_{-\tilde{v}_1}\O\rangle.
	\end{aligned}
\]
We now have to assume the existence of $i,j=1,\dots,m$ such that $p+r=-v_i$ and $q+r=-\tilde{v}_j$ otherwise the expectation on the last line would vanish. In  particular we assume $i=j=m$ since the $m^2$ cases are all equivalent. Pairing the remaining momenta in $P_S$ we obtain
\[
	\begin{aligned}
		\langle\xi_\nu,&\cE_{M,1}\xi_\nu\rangle\\
		=&\sum_{m\geq 1}\frac1{(m-1)!}\frac{1}{N^{m+1}}\hspace{-0.2cm}\sum_{\substack{v_1,\tilde{v}_1\in P_S,r_1,\tilde{r}_1\in P_H:\\r_1+v_1,\tilde{r}_1+\tilde{v}_1\in P_H}}\hspace{-0.2cm}\cdots\hspace{-0.1cm}\sum_{\substack{v_m,\tilde{v}_m\in P_S,r_m,\tilde{r}_m\in P_H:\\r_m+v_m,\tilde{r}_m+\tilde{v}_m\in P_H}}\hspace{-0.2cm}\theta\big(\{r_j,v_j\}_{j=1}^m\big)\theta\big(\{\tilde{r}_j,\tilde{v}_j\}_{j=1}^m\big)\\
		&\times\prod_{j=1}^m\eta_{r_j}\eta_{\tilde{r}_j}\s_{v_j}\s_{\tilde{v}_j}\prod_{i=1}^{m-1}\d_{v_i,\tilde{v}_i}\sum_{\substack{r\in \L^*:\\r+v_m,r+\tilde{v}_m\in P_H}}\varphi_1(-r-v_m,-r-\tilde{v}_m,r)\\
		&\times\langle \O, A_{r_m,v_m}\cdots A_{r_1,v_1}a^*_{-r-\tilde{v}_m}a_{-r-v_m}A^*_{\tilde{r}_m,\tilde{v}_m} \cdots A^*_{\tilde{r}_1,v_1}\O\rangle.
	\end{aligned}
\]
At this point we recognize that, due to the presence of the restrictions encoded in $\theta\big(\{r_j,v_j\}_{j=1}^m\big)\theta\big(\{\tilde{r}_j,\tilde{v}_j\}_{j=1}^m\big)$, if the operator $a_{-r-v_m}$ is annihilated with an operator in $A^*_{\tilde{r}_j,v_j}$ then the operator $a^*_{-r-\tilde{v}_m}$ has to be annihilated with an operator in $A_{r_j,v_j}$. In particular, if $j=m$ then the remaining operator in $A^*_{\tilde{r}_m,\tilde v_m}$ has to be contracted with the remaining operator in $A_{r_m,v_m}$ and we obtain a contribution denoted by $M_1.$ On the other hand, if $j\neq m$ (by symmetry we assume $j=m-1$) we distinguish two cases: either the remaining operator in $A^*_{\tilde{r}_{m-1},v_{m-1}}$ is annihilated with the remaining operator in $A_{r_{m-1},v_{m-1}}$ and the operators $A^*_{\tilde{r}_m,\tilde{v}_m}$ and $A_{r_m,v_m}$ are contracted among themselves (imposing $v_m=\tilde{v}_m$) or the remaining operator in $A^*_{\tilde{r}_{m-1},v_{m-1}}$ is contracted with an operator in $A_{r_m,v_m}$ and the remaining operator in $A_{r_{m-1},v_{m-1}}$ is contracted with an operator in $A^*_{\tilde{r}_m,\tilde{v}_m}$ (then we are left with one operator in $A^*_{\tilde{r}_m,\tilde{v}_m}$ and one in $A_{r_m,v_m}$ that are necessarily contracted with each other). We denote these contributions with $M_2$ and $M_3$ respectively.\\
 Explicitly,
 \be\label{eq:cEM1split}
 	\langle\xi_\nu,\cE_{M,1}\xi_\nu\rangle=M_1+M_2+M_3
 \ee
 with
\[
	\begin{aligned}
		M_1=&\sum_{m\geq 1}\frac1{(m-1)!}\frac{1}{N^{m+1}}\hspace{-0.2cm}\sum_{\substack{v_1,\tilde{v}_1\in P_S,r_1,\tilde{r}_1\in P_H:\\r_1+v_1,\tilde{r}_1+\tilde{v}_1\in P_H}}\hspace{-0.2cm}\cdots\hspace{-0.1cm}\sum_{\substack{v_m,\tilde{v}_m\in P_S,r_m,\tilde{r}_m\in P_H:\\r_m+v_m,\tilde{r}_m+\tilde{v}_m\in P_H}}\hspace{-0.2cm}\theta\big(\{r_j,v_j\}_{j=1}^m\big)\theta\big(\{\tilde{r}_j,\tilde{v}_j\}_{j=1}^m\big)\\
		&\times\prod_{j=1}^m\eta_{r_j}\eta_{\tilde{r}_j}\s_{v_j}\s_{\tilde{v}_j}\prod_{i=1}^{m-1}\d_{v_i,\tilde{v}_i}(\d_{\tilde{r}_i}+\d_{\tilde{r}_i+v_i})\sum_{r\in \L^*}\varphi_1(-r-v_m,-r-\tilde{v}_m,r)\\
		&\times\sum_{\substack{p_m\in \{-r_m,r_m+v_m\}\\\tilde{p}_m\in \{\tilde{r}_m,\tilde{r}_m+\tilde{v}_m\}}}\d_{r,-p_m-\tilde{v}_m}\d_{\tilde{p}_m,p_m+\tilde{v}_m-v_m}\\
		M_2=&\sum_{m\geq 2}\frac1{(m-2)!}\frac{1}{N^{m+1}}\hspace{-0.2cm}\sum_{\substack{v_1,\tilde{v}_1\in P_S,r_1,\tilde{r}_1\in P_H:\\r_1+v_1,\tilde{r}_1+\tilde{v}_1\in P_H}}\hspace{-0.2cm}\cdots\hspace{-0.1cm}\sum_{\substack{v_m,\tilde{v}_m\in P_S,r_m,\tilde{r}_m\in P_H:\\r_m+v_m,\tilde{r}_m+\tilde{v}_m\in P_H}}\hspace{-0.2cm}\theta\big(\{r_j,v_j\}_{j=1}^m\big)\theta\big(\{\tilde{r}_j,\tilde{v}_j\}_{j=1}^m\big)\\
		&\times\prod_{j=1}^m\eta_{r_j}\eta_{\tilde{r}_j}\s_{v_j}\s_{\tilde{v}_j}\prod_{\substack{i=1\\i\neq m-1}}^{m}\d_{v_i,\tilde{v}_i}(\d_{r_i,\tilde{r}_i}+\d_{-r_i,\tilde{r}_i+v_i})\sum_{r\in \L^*}\varphi_1(-r-v_m,-r-{v}_m,r)\\
		&\times\d_{\tilde{v}_{m-1},v_{m-1}}\sum_{\substack{p_{m-1}\in \{-r_{m-1},r_{m-1}+v_{m-1}\}\\\tilde p_{m-1}\in\{-\tilde{r}_{m-1},\tilde{r}_{m-1}+v_{m-1}\}}}\d_{r,-p_{m-1}-{v}_m}\d_{\tilde{p}_{m-1},p_{m-1}}\\
		M_3=&\sum_{m\geq 2}\frac1{(m-2)!}\frac{1}{N^{m+1}}\hspace{-0.2cm}\sum_{\substack{v_1,\tilde{v}_1\in P_S,r_1,\tilde{r}_1\in P_H:\\r_1+v_1,\tilde{r}_1+\tilde{v}_1\in P_H}}\hspace{-0.2cm}\cdots\hspace{-0.1cm}\sum_{\substack{v_m,\tilde{v}_m\in P_S,r_m,\tilde{r}_m\in P_H:\\r_m+v_m,\tilde{r}_m+\tilde{v}_m\in P_H}}\hspace{-0.2cm}\theta\big(\{r_j,v_j\}_{j=1}^m\big)\theta\big(\{\tilde{r}_j,\tilde{v}_j\}_{j=1}^m\big)\\
		&\times\prod_{j=1}^m\eta_{r_j}\eta_{\tilde{r}_j}\s_{v_j}\s_{\tilde{v}_j}\prod_{i=1}^{m-2}\d_{v_i,\tilde{v}_i}(\d_{\tilde{r}_i}+\d_{\tilde{r}_i+v_i})\d_{\tilde v_{m-1},v_{m-1}}\sum_{r\in \L^*}\varphi_1(-r-v_m,-r-\tilde{v}_m,r)\\
		&\times\hspace{-0.3cm}\sum_{\substack{p_\ell\in \{-r_\ell,r_\ell+v_\ell\}\\\tilde{p}_\ell\in \{\tilde{r}_\ell,\tilde{r}_\ell+\tilde{v}_\ell\}\\\ell=m-1,m}}\hspace{-0.3cm}\d_{r,-p_{m-1}-\tilde{v}_m}\d_{\tilde{p}_{m-1},p_{m-1}+\tilde{v}_m-v_m}\d_{\tilde p_m,p_m-v_m+\tilde v_m}\d_{p_{m-1},-p_m+v_m-\tilde v_m+v_{m-1}}.
	\end{aligned}
\]
Since,
\[
	\begin{aligned}
		|\varphi_1(-r-v_m,-r-&\tilde v_m,r)\d_{r,-p_m-\tilde{v}_m}\d_{\tilde{p}_m,p_m+\tilde{v}_m-v_m}|\\
		&\leq CN^\k (\|\eta_H\|_\infty^2|\s_{v_m}||\s_{\tilde v_m}|+|\g_{v_m}||\g_{\tilde v_m}|+\|\eta_H\|_\infty|\g_{v_m}||\s_{\tilde v_m}|)
	\end{aligned}
\]
we have
\be\label{eq:M1}
	\begin{aligned}
		\frac{|M_1|}{\|\xi_\nu\|^2}\leq& CN^{\k-2}\|\eta_H\|^2(\|\s_S\|^4\|\eta_H\|_\infty^2+\|\s_S\g_S\|_1^2+\|\s_S\|^2\|\g_S\s_S\|_1\|\eta_H\|_\infty)
	\end{aligned}
\ee
On the other hand, using the fact that 
\[
	\begin{aligned}
		|\varphi_1(-r-v_m,&-r-{v}_m,r)|\d_{r,-p_{m-1}-{v}_m}\d_{\tilde{p}_{m-1},p_{m-1}}|\\
		&\leq CN^\k (\|\eta_H\|_\infty^2\|\s_S\|_\infty^2+\|\g_S\|_\infty^2+\|\g_S\|_\infty\|\s_S\|_\infty\|\eta_H\|_\infty)
	\end{aligned}
\]
we obtain
\be\label{eq:M2}
	\begin{aligned}
		\frac{|M_2|}{\|\xi_\nu\|^2}\leq&CN^{\k-3}\|\eta_H\|^4\|\s_S\|^4(\|\s_S\|_{\infty}^2\|\eta_H\|_\infty^2+\|\g_S\|_\infty^2+\|\g_S\|_\infty\|\s_S\|_\infty\|\eta_H\|_\infty)\\
	\end{aligned}
\ee
Finally, we note that
\[
	\begin{aligned}
		|\varphi_1(-r-v_m,&-r-\tilde{v}_m,r)\d_{r,-p_{m-1}-\tilde{v}_m}\d_{\tilde{p}_{m-1},p_{m-1}+\tilde{v}_m-v_m}|\\
		&\leq CN^\k (\|\eta_H\|_\infty^2|\s_{v_m}||\s_{\tilde{v}_m}|+|\g_{v_m}||\g_{\tilde{v}_m}|+\|\eta_H\|_\infty|\g_{v_m}||\s_{\tilde{v}_m}|).
	\end{aligned}
\]
Therefore,
\be\label{eq:M3}
	\begin{aligned}
		\frac{|M_3|}{\|\xi_\nu\|^2}\leq& N^{\k-3}\|\eta_H\|^2\|\eta_H\|_\infty^2\|\s_S\|^2(\|\s_S\|^4\|\eta_H\|_\infty^2+\|\g_S\s_S\|_1^2+\|\s_S\|^2\|\g_S\s_S\|_1\|\eta_H\|_\infty).
	\end{aligned}
\ee
By \eqref{eq:cEM1split},\eqref{eq:M1},\eqref{eq:M2},\eqref{eq:M3} we conclude, using \eqref{eq:eta_H} and \eqref{eq:tau_S}, that for any $\k<2/3$ and $\e$ small enough
\be\label{eq:cEM1}
	\begin{aligned}
		\frac{\langle\xi_\nu,\cE_{M,1}\xi_\nu\rangle}{\|\xi_\nu\|^2}\leq& CN^{5\k/2-\e}.
	\end{aligned}
\ee
We now consider the expectation on $\xi_\nu$ of $\cE_{M,2}.$ We have by definition
\[
	\begin{aligned}
		\langle\xi_\nu&,\cE_{M,2}\xi_\nu\rangle\\
		=&\sum_{m\geq 1}\frac1{(m!)^2}\frac{1}{N^{m+1}}\hspace{-0.2cm}\sum_{\substack{v_1,\tilde{v}_1\in P_S,r_1,\tilde{r}_1\in P_H:\\r_1+v_1,\tilde{r}_1+\tilde{v}_1\in P_H}}\hspace{-0.2cm}\cdots\hspace{-0.1cm}\sum_{\substack{v_m,\tilde{v}_m\in P_S,r_m,\tilde{r}_m\in P_H:\\r_m+v_m,\tilde{r}_m+\tilde{v}_m\in P_H}}\hspace{-0.2cm}\theta\big(\{r_j,v_j\}_{j=1}^m\big)\theta\big(\{\tilde{r}_j,\tilde{v}_j\}_{j=1}^m\big)\\
		&\times\prod_{j=1}^m\eta_{r_j}\eta_{\tilde{r}_j}\s_{v_j}\s_{\tilde{v}_j}\sum_{r\in \L^*}\sum_{\substack{p,q\in P_H:\\p+r,q+r\in P_S}}\varphi_2(p,q,r)\\
		&\times\langle \O, a_{r_m+v_m}\cdots a_{-v_1}a^*_{p+r}a^*_{-p}a_{q+r}a_{-q}a^*_{\tilde{r}_m+\tilde{v}_m} \cdots a^*_{-\tilde{v}_1}\O\rangle.
	\end{aligned}
\]
We note that there are necessarily $i,j$ such that $p+r=-v_i$ and $q+r=-\tilde v_j.$ Assuming by symmetry $i=j=m$ and pairing the remaining $m-1$ momenta in $P_S $ we get
\[
	\begin{aligned}
		\langle\xi_\nu&,\cE_{M,2}\xi_\nu\rangle\\
			=&\sum_{m\geq 1}\frac1{(m-1)!}\frac{1}{N^{m+1}}\hspace{-0.2cm}\sum_{\substack{v_1,\tilde{v}_1\in P_S,r_1,\tilde{r}_1\in P_H:\\r_1+v_1,\tilde{r}_1+\tilde{v}_1\in P_H}}\hspace{-0.2cm}\cdots\hspace{-0.1cm}\sum_{\substack{v_m,\tilde{v}_m\in P_S,r_m,\tilde{r}_m\in P_H:\\r_m+v_m,\tilde{r}_m+\tilde{v}_m\in P_H}}\hspace{-0.2cm}\theta\big(\{r_j,v_j\}_{j=1}^m\big)\theta\big(\{\tilde{r}_j,\tilde{v}_j\}_{j=1}^m\big)\\
		&\times\prod_{j=1}^m\eta_{r_j}\eta_{\tilde{r}_j}\s_{v_j}\s_{\tilde{v}_j}\prod_{i=1}^{m-1}\d_{\tilde{v}_i,v_i}\sum_{\substack{r\in \L^*:\\r+v_m,r+\tilde v_m\in P_H}}\varphi_2(-r-v_m,-r-\tilde v_m,r)\\
		&\times\langle \O, A_{r_m,v_m}\cdots A_{r_1,v_1}a^*_{r+v_m}a_{r+\tilde v_m}A^*_{\tilde{r}_m,\tilde v_m}A^*_{\tilde r_{m-1},v_{m-1}} \cdots A^*_{\tilde r_1,v_1}\O\rangle\\
		=&\widetilde{M}_1+\widetilde{M}_2+\widetilde{M}_3
	\end{aligned}
\]
where
\[
	\begin{aligned}
		\widetilde{M}_1=&\sum_{m\geq 1}\frac1{(m-1)!}\frac{1}{N^{m+1}}\hspace{-0.2cm}\sum_{\substack{v_1,\tilde{v}_1\in P_S,r_1,\tilde{r}_1\in P_H:\\r_1+v_1,\tilde{r}_1+\tilde{v}_1\in P_H}}\hspace{-0.2cm}\cdots\hspace{-0.1cm}\sum_{\substack{v_m,\tilde{v}_m\in P_S,r_m,\tilde{r}_m\in P_H:\\r_m+v_m,\tilde{r}_m+\tilde{v}_m\in P_H}}\hspace{-0.2cm}\theta\big(\{r_j,v_j\}_{j=1}^m\big)\theta\big(\{\tilde{r}_j,\tilde{v}_j\}_{j=1}^m\big)\\
		&\times\prod_{j=1}^m\eta_{r_j}\eta_{\tilde{r}_j}\s_{v_j}\s_{\tilde{v}_j}\prod_{i=1}^{m-1}\d_{v_i,\tilde{v}_i}(\d_{\tilde{r}_i}+\d_{\tilde{r}_i+v_i})\sum_{r\in \L^*}\varphi_2(-r-v_m,-r-\tilde{v}_m,r)\\
		&\times\sum_{\substack{p_m\in \{-r_m,r_m+v_m\}\\\tilde{p}_m\in \{\tilde{r}_m,\tilde{r}_m+\tilde{v}_m\}}}\d_{r,p_m-{v}_m}\d_{\tilde{p}_m,p_m+\tilde{v}_m-v_m}\\
		\widetilde{M}_2=&\sum_{m\geq 2}\frac1{(m-2)!}\frac{1}{N^{m+1}}\hspace{-0.2cm}\sum_{\substack{v_1,\tilde{v}_1\in P_S,r_1,\tilde{r}_1\in P_H:\\r_1+v_1,\tilde{r}_1+\tilde{v}_1\in P_H}}\hspace{-0.2cm}\cdots\hspace{-0.1cm}\sum_{\substack{v_m,\tilde{v}_m\in P_S,r_m,\tilde{r}_m\in P_H:\\r_m+v_m,\tilde{r}_m+\tilde{v}_m\in P_H}}\hspace{-0.2cm}\theta\big(\{r_j,v_j\}_{j=1}^m\big)\theta\big(\{\tilde{r}_j,\tilde{v}_j\}_{j=1}^m\big)\\
		&\times\prod_{j=1}^m\eta_{r_j}\eta_{\tilde{r}_j}\s_{v_j}\s_{\tilde{v}_j}\prod_{\substack{i=1\\i\neq m-1}}^{m}\d_{v_i,\tilde{v}_i}(\d_{r_i,\tilde{r}_i}+\d_{-r_i,\tilde{r}_i+v_i})\sum_{r\in \L^*}\varphi_2(-r-v_m,-r-{v}_m,r)\\
		&\times\d_{\tilde{v}_{m-1},v_{m-1}}\sum_{\substack{p_{m-1}\in \{-r_{m-1},r_{m-1}+v_{m-1}\}\\\tilde p_{m-1}\in\{-\tilde{r}_{m-1},\tilde{r}_{m-1}+v_{m-1}\}}}\d_{r,-p_{m-1}-{v}_m}\d_{\tilde{p}_{m-1},p_{m-1}}\\
		\widetilde M_3=&\sum_{m\geq 2}\frac1{(m-2)!}\frac{1}{N^{m+1}}\hspace{-0.2cm}\sum_{\substack{v_1,\tilde{v}_1\in P_S,r_1,\tilde{r}_1\in P_H:\\r_1+v_1,\tilde{r}_1+\tilde{v}_1\in P_H}}\hspace{-0.2cm}\cdots\hspace{-0.1cm}\sum_{\substack{v_m,\tilde{v}_m\in P_S,r_m,\tilde{r}_m\in P_H:\\r_m+v_m,\tilde{r}_m+\tilde{v}_m\in P_H}}\hspace{-0.2cm}\theta\big(\{r_j,v_j\}_{j=1}^m\big)\theta\big(\{\tilde{r}_j,\tilde{v}_j\}_{j=1}^m\big)\\
		&\times\prod_{j=1}^m\eta_{r_j}\eta_{\tilde{r}_j}\s_{v_j}\s_{\tilde{v}_j}\prod_{i=1}^{m-2}\d_{v_i,\tilde{v}_i}(\d_{\tilde{r}_i}+\d_{\tilde{r}_i+v_i})\d_{\tilde v_{m-1},v_{m-1}}\sum_{r\in \L^*}\varphi_2(-r-v_m,-r-\tilde{v}_m,r)\\
		&\times\sum_{\substack{p_\ell\in \{-r_\ell,r_\ell+v_\ell\}\\\tilde{p}_\ell\in \{\tilde{r}_\ell,\tilde{r}_\ell+\tilde{v}_\ell\}\\\ell=m-1,m}}\d_{r,p_{m-1}-{v}_m}\d_{\tilde{p}_{m-1},p_{m-1}+\tilde{v}_m-v_m}\d_{\tilde p_m,p_m-v_m+\tilde v_m}\d_{p_{m-1},-p_m+v_m-\tilde v_m+v_{m-1}}.
	\end{aligned}
\]
Thus, the bound
\[
	\begin{aligned}
	|\varphi_2(&-r-v_m,-r-\tilde{v}_m,r)\d_{r,p_m-{v}_m}\d_{\tilde{p}_m,p_m+\tilde{v}_m-v_m}|\\
	&\leq CN^\k(\|\eta_H\|_\infty^2|\g_{v_m}||\g_{\tilde v_m}|+\|\eta_H\|_\infty|\g_{\tilde v_m}||\s_{v_m}|+\|\eta_H\|_\infty|\g_{v_m}||\s_{\tilde v_m}|+|\s_{v_m}||\s_{\tilde{v}_m}|)
	\end{aligned}
\]
implies
\be\label{eq:M1tilde}
	\begin{aligned}
		\frac{|\widetilde{M}_1|}{\|\xi_\nu\|^2}\leq& CN^{\k-2}\|\eta_H\|^2(\|\g_S\s_S\|_1^2\|\eta_H\|_\infty^2+\|\g_S\s_S\|_1\|\s_S\|^2\|\eta_H\|_\infty+\|\s_S\|^4).
	\end{aligned}
\ee
On the other hand, noting that
\[
	\begin{aligned}
		|\varphi_2(-r-v_m,-r-{v}_m,r)&\d_{\tilde{v}_{m-1},v_{m-1}}\d_{r,-p_{m-1}-{v}_m}\d_{\tilde{p}_{m-1},p_{m-1}}|\\
		&\leq CN^\k(\|\g_S\|_\infty^2\|\eta_H\|_\infty^2+\|\g_S\|_\infty\|\s_S\|_\infty\|\eta_H\|_\infty+\|\s_S\|_\infty^2)\\
	\end{aligned}
\]
we get
\[
	\begin{aligned}
		\frac{|\widetilde M_2|}{\|\xi_\nu\|^2}\leq&CN^{\k-3}\|\eta_H\|^4\|\s_S\|^4(\|\g_S\|_\infty^2\|\eta_H\|_\infty^2+\|\eta_H\|_\infty\|\s_S\|_\infty\|\g_S\|_\infty+\|\s_S\|_\infty^2).
	\end{aligned}
\]
Finally, by
\[
	\begin{aligned}
		&|\varphi_2(-r-v_m,-r-\tilde{v}_m,r)\d_{r,p_{m-1}-{v}_m}\d_{\tilde{p}_{m-1},p_{m-1}+\tilde{v}_m-v_m}|\\
		&\leq CN^\k(\|\eta_H\|_\infty^2|\g_{v_m}||\g_{\tilde v_m}|+\|\eta_H\|_\infty|\g_{\tilde v_m}||\s_{v_m}|+\|\eta_H\|_\infty|\g_{v_m}||\s_{\tilde v_m}|+|\s_{v_m}||\s_{\tilde v_m}|)
	\end{aligned}
\]
we obtain
\be\label{eq:M3tilde}
	\begin{aligned}
		\frac{|\widetilde M_3|}{\|\xi_\nu\|^2}\leq & CN^{\k-3}\|\eta_H\|^2\|\eta_H\|_\infty^2\|\s_S\|^2(\|\g_S\s_S\|_1^2\|\eta_H\|_\infty^2+\|\g_S\s_S\|_1\|\s_S\|^2\|\eta_H\|_\infty+\|\s_S\|^4).
	\end{aligned}
\ee
From \eqref{eq:M1tilde}-\eqref{eq:M3tilde} and recalling \eqref{eq:eta_H} and \eqref{eq:tau_S}, we obtain for $\k<2/3$ and $\e$ small enough
\be\label{eq:cEM2}
	\begin{aligned}
		&\frac{\langle\xi_\nu,\cE_{M,2}\xi_\nu\rangle}{\|\xi_\nu\|^2}\leq \frac{|\widetilde M_1|+|\widetilde M_2|+|\widetilde M_3|}{\|\xi_\nu\|^2}\leq CN^{5\k/2-\e}
	\end{aligned}
\ee
To conclude we still have to bound the expectation of $\cE_{M,3}$ on the state $\xi_\nu$. 
Proceeding similarly as before we write
\[
		\langle\xi_\nu,\cE_{M,3}\xi_\nu\rangle
		=M_1'+M_2'+M_3'
\]
where we introduced the notations
\[
	\begin{aligned}
		M_1'=&\sum_{m\geq1}\frac1{(m-1)!}\frac1{N^{m+1}}\hspace{-0.2cm}\sum_{\substack{v_1,\tilde{v}_1\in P_S,r_1,\tilde{r}_1\in P_H:\\r_1+v_1,\tilde{r}_1+\tilde{v}_1\in P_H}}\hspace{-0.2cm}\cdots\hspace{-0.1cm}\sum_{\substack{v_m,\tilde{v}_m\in P_S,r_m,\tilde{r}_m\in P_H:\\r_m+v_m,\tilde{r}_m+\tilde{v}_m\in P_H}}\hspace{-0.2cm}\theta\big(\{r_j,v_j\}_{j=1}^m\big)\theta\big(\{\tilde{r}_j,\tilde{v}_j\}_{j=1}^m\big)\\
		&\times\prod_{j=1}^m\eta_{r_j}\eta_{\tilde{r}_j}\s_{v_j}\s_{\tilde{v}_j}\prod_{i=1}^{m-1}\d_{v_i,\tilde{v}_i}(\d_{\tilde r_i,r_i}+\d_{-\tilde r_i,r_i+v_i})\sum_{q\in P_H}\varphi_3(-v_m,q,v_m-\tilde v_m)\\
		&\times\sum_{\substack{p_m\in \{-r_m,r_m+v_m\}\\\tilde p_m\in \{-\tilde r_m,\tilde r_m+\tilde v_m\}}}\d_{\tilde p_m,q}\d_{q,p_m+\tilde v_m-v_m}\\
		M_2'=&\sum_{m\geq 2}\frac1{(m-2)!}\frac{1}{N^{m+1}}\hspace{-0.2cm}\sum_{\substack{v_1,\tilde{v}_1\in P_S,r_1,\tilde{r}_1\in P_H:\\r_1+v_1,\tilde{r}_1+\tilde{v}_1\in P_H}}\hspace{-0.2cm}\cdots\hspace{-0.1cm}\sum_{\substack{v_m,\tilde{v}_m\in P_S,r_m,\tilde{r}_m\in P_H:\\r_m+v_m,\tilde{r}_m+\tilde{v}_m\in P_H}}\hspace{-0.2cm}\theta\big(\{r_j,v_j\}_{j=1}^m\big)\theta\big(\{\tilde{r}_j,\tilde{v}_j\}_{j=1}^m\big)\\
		&\times\prod_{j=1}^m\eta_{r_j}\eta_{\tilde{r}_j}\s_{v_j}\s_{\tilde{v}_j}\prod_{\substack{i=1\\i\neq m-1}}^{m}\d_{v_i,\tilde{v}_i}(\d_{r_i,\tilde{r}_i}+\d_{-r_i,\tilde{r}_i+v_i})\sum_{q\in P_H}\varphi_3(-v_m,q,v_m-\tilde v_m)\\
		&\times\d_{\tilde{v}_{m-1},v_{m-1}}\sum_{\substack{p_{m-1}\in \{-r_{m-1},r_{m-1}+v_{m-1}\}\\\tilde p_{m-1}\in\{-\tilde{r}_{m-1},\tilde{r}_{m-1}+v_{m-1}\}}}\d_{q,p_{m-1}}\d_{\tilde{p}_{m-1},p_{m-1}}
	\end{aligned}
\]
\[
	\begin{aligned}
		M_3'=&\sum_{m\geq 2}\frac1{(m-2)!}\frac{1}{N^{m+1}}\hspace{-0.2cm}\sum_{\substack{v_1,\tilde{v}_1\in P_S,r_1,\tilde{r}_1\in P_H:\\r_1+v_1,\tilde{r}_1+\tilde{v}_1\in P_H}}\hspace{-0.2cm}\cdots\hspace{-0.1cm}\sum_{\substack{v_m,\tilde{v}_m\in P_S,r_m,\tilde{r}_m\in P_H:\\r_m+v_m,\tilde{r}_m+\tilde{v}_m\in P_H}}\hspace{-0.2cm}\theta\big(\{r_j,v_j\}_{j=1}^m\big)\theta\big(\{\tilde{r}_j,\tilde{v}_j\}_{j=1}^m\big)\\
		&\times\prod_{j=1}^m\eta_{r_j}\eta_{\tilde{r}_j}\s_{v_j}\s_{\tilde{v}_j}\prod_{i=1}^{m-2}\d_{v_i,\tilde{v}_i}(\d_{\tilde{r}_i}+\d_{\tilde{r}_i+v_i})\d_{\tilde v_{m-1},v_{m-1}}\sum_{q\in P_H}\varphi_3(-v_m,q,v_m-\tilde v_m)\\
		&\times\sum_{\substack{p_\ell\in \{-r_\ell,r_\ell+v_\ell\}\\\tilde{p}_\ell\in \{\tilde{r}_\ell,\tilde{r}_\ell+\tilde{v}_\ell\}\\\ell=m-1,m}}\d_{q,p_{m-1}+\tilde{v}_m-v_m}\d_{\tilde{p}_{m-1},q}\d_{\tilde p_m,p_m-v_m+\tilde v_m}\d_{p_{m-1},-p_m+v_m-\tilde v_m+v_{m-1}}.
	\end{aligned}
\]
Hence, we have
\be\label{eq:M1'}
	\begin{aligned}
		\frac{|M_1'|}{\|\xi_\nu\|^2}\leq& CN^{\k-1}\|\eta_H\|^2\\
		&\hspace{1cm}\times(\|\s_S\|^4\|\eta_H\|_\infty^2+\|\g_S\s_S\|_1^2+\|\g_S\s_S\|_1^2\|\eta_H\|_\infty^2+\|\s_S\|^2\|\g_S\s_S\|_1\|\eta_H\|_\infty)\\
	\end{aligned}
\ee
where we used the bound
\[
	\begin{aligned}
		&|\varphi_3(-v_m,q,v_m-\tilde v_m)\d_{\tilde p_m,q}\d_{q,p_m+\tilde v_m-v_m}|\\
		&\leq CN^\k(\|\eta_H\|_\infty^2|\s_{v_m}||\s_{\tilde v_m}|+|\g_{v_m}||\g_{\tilde v_m}|+\|\eta_H\|_\infty^2|\g_{v_m}||\g_{\tilde v_m}|+\|\eta_H\|_\infty|\s_{\tilde v_m}||\g_{v_m}|).
	\end{aligned}
\]
Moreover, since
\[
	\begin{aligned}
		&|\varphi_3(-v_m,q,v_m-\tilde v_m)\d_{q,p_{m-1}}\d_{\tilde{p}_{m-1},p_{m-1}}|\\
		&\leq CN^\k(\|\eta_H\|_\infty^2\|\s_S\|_\infty^2+\|\g_S\|_\infty^2+\|\g_S\|_\infty^2\|\|\eta_H\|_\infty^2+\|\s_S\|_\infty\|\eta_H\|_\infty\|\g_S\|_\infty)
	\end{aligned}
\]
we get
\[
	\begin{aligned}
		\frac{|M_2'|}{\|\xi_\nu\|^2}\leq &CN^{\k-3}\|\eta_H\|^4\|\s_S\|^4\\
		&\hspace{1cm}\times(\|\s_S\|_\infty^2\|\eta_H\|_\infty^2+\|\g_S\|_\infty^2+\|\g_S\|_\infty^2\|\eta_H\|_\infty^2+\|\g_S\|_\infty\|\s_S\|_\infty\|\eta_H\|_\infty).
	\end{aligned}
\]
Finally, we note that
\[
	\begin{aligned}
		&|\varphi_3(-v_m,q,v_m-\tilde v_m)\d_{q,p_{m-1}+\tilde{v}_m-v_m}\d_{\tilde{p}_{m-1},q}|\\
		&\leq CN^\k(\|\eta_H\|_\infty^2\|\s_{v_m}||\s_{\tilde v_m}|+|\g_{v_m}||\g_{\tilde v_m}|+\|\eta_H\|_\infty^2|\g_{v_m}||\g_{\tilde v_m}|+\|\eta_H\|_\infty\|\s_{\tilde v_m}||\g_{v_m}|)
	\end{aligned}
\]
which implies
\be\label{eq:M3'}
	\begin{aligned}
		\frac{|M_3'|}{\|\xi_\nu\|^2}\leq &CN^{\k-3}\|\s_S\|^2\|\eta_H\|^2\|\eta_H\|_\infty^2(\|\s_S\|^4\|\eta_H\|_\infty^2+\|\g_S\s_S\|^2+\|\g_S\s_S\|_1\|\s_S\|^2\|\eta_H\|_\infty).
	\end{aligned}
\ee
The bounds \eqref{eq:M1'}-\eqref{eq:M3'} yield, recalling \eqref{eq:eta_H} and \eqref{eq:tau_S},
\be\label{eq:cEM3}
	\begin{aligned}
		&\frac{\langle\xi_\nu,\cE_{M,3}\xi_\nu\rangle}{\|\xi_\nu\|^2}\leq\frac{|M_1'|+|M_2'|+|M_3'|}{\|\xi_\nu\|^2}\leq CN^{5\k/2-\e}
	\end{aligned}
\ee
under the assumptions $\k<2/3$ and $\e$ small enough.\\
From \eqref{eq:cE2},\eqref{eq:cEC},\eqref{eq:cEH},\eqref{eq:cES},\eqref{eq:cEM1},\eqref{eq:cEM2} and \eqref{eq:cEM3} we get \eqref{eq:last}.


\begin{thebibliography}{55}
\bibitem{ABS}A. Adhikari, C. Brennecke, B. Schlein. Bose-Einstein Condensation Beyond the Gross-Pitaevskii Regime. {\it Ann. Henri Poincar\'e} {\bf22} (2021), 1163-1233.
\bibitem{Ba}
T. Balaban, J. Feldman, H. Knoerrer, E. Trubowitz. Complex Bosonic Many-Body Models: Overview of the Small Field Parabolic Flow.  {\it Ann. Henri Poincar\'e} {\bf 18} (2017), 2873--2903.
\bibitem{BCOPS} G. Basti, S. Cenatiempo, A. Olgiati, G. Pasqualetti, B. Schlein. A second order upper bound for the energy of hard core bosons in the Gross-Pitaevskii limit. In preparation.
\bibitem{BCOPSII} G. Basti, S. Cenatiempo, A. Olgiati, G. Pasqualetti, B. Schlein. Ground state energy of a Bose gas in the Gross-Pitaevskii regime. Preprint  	arXiv:2202.10270
\bibitem{BCS}
		G. Basti, S. Cenatiempo, B. Schlein. A new second order upper bound for the ground state energy of dilute Bose gases. {\it Forum Math. Sigma} {\bf 9} (2021), Paper No. e74.
\bibitem{Be}
G. Benfatto. Renormalization group approach to zero temperature Bose condensation. In: Proceedings of the workshop “Constructive results in Field Theory, Statistical Mechanics and Condensed Matter Physics”, Palaiseau, July 25–27, pp. 219–247 (1994)
\bibitem{BBCS}
C. Boccato, C. Brennecke, S. Cenatiempo, B. Schlein. Optimal rate for Bose- Einstein condensation in the Gross-Pitaevskii regime. {\it Commun. Math. Phys} {\bf376} (2020), 1311–1395 .
\bibitem{BBCSacta}C. Boccato, C. Brennecke, S. Cenatiempo, B. Schlein. Bogoliubov Theory in the Gross-Pitaevskii Limit. {\it Acta Mathematica} {\bf 222} (2019), no. 2, 219335.
\bibitem{Bog}
N. N. Bogoliubov. On the theory of superfluidity.
{\it Izv. Akad. Nauk. USSR} {\bf 11} (1947), 77. Engl. Transl. {\it J. Phys. (USSR)} {\bf 11} (1947), 23.     
\bibitem{BrCS} C. Brennecke, M. Caporaletti, B. Schlein. Excitation Spectrum for Bose Gases beyond the Gross-Pitaevskii Regime. Preprint: arXiv:2104.13003.
\bibitem{DG}
 D. Dimonte, E. L. Giacomelli. On Bose-Einstein condensates in the Thomas-Fermi regime. Preprint: arXiv:2112.02343.
\bibitem{D}
F.J. Dyson. Ground-State Energy of a Hard-Sphere Gas. {\it Phys. Rev.} {\bf 106} (1957), 20--26.  
	\bibitem{ESY}
L. Erd\H os, B. Schlein, H.-T. Yau. Ground-state energy of a low-density Bose gas: a second order upper bound. {\it Phys. Rev. A} {\bf 78} (2008), 053627. 
\bibitem{FGHP}
M. Falconi, E. L. Giacomelli, C. Hainzl, M. Porta. The Dilute Fermi Gas via Bogoliubov Theory. {\it Ann. Henri Poincar\'e} {\bf22},  (2021), 2283--2353.
\bibitem{F}S. Fournais. Length scales for BEC in the dilute Bose gas. Preprint arXiv:2011.00309. 
\bibitem{FS}
S. Fournais, J.P. Solovej. The energy of dilute Bose gases. {\it Ann. Math.} {\bf 192}(3) (2020), 893--976.  
\bibitem{FSII} S. Fournais, J.P. Solovej. The energy of dilute Bose gases II:
The general case. Preprint arXiv: 2108.12022. 
\bibitem{HY}
K. Huang and C. N. Yang. Quantum-Mechanical Many-Body Problem with Hard-Sphere Interaction. {\it Phys. Rev.} {\bf 105}, 767-775 (1957).
\bibitem{LHY}
T. D. Lee, K. Huang, and C. N. Yang. Eigenvalues and eigenfunctions of a Bose
system of hard spheres and its low-temperature properties. {\it Physical Review}, {\bf 106} (1957), 1135--1145. 
\bibitem{LSS}
E. H. Lieb, R. Seiringer, J. P. Solovej. Ground-state energy of the low-density Fermi gas. {\it Phys. Rev. A} {\bf 71} (2005)
\bibitem{LSY}
E. H. Lieb, R. Seiringer, and J. Yngvason. Bosons in a trap: A rigorous derivation of the Gross-Pitaevskii energy functional. {\it Phys. Rev. A} {\bf 61} (2000), 043602.
\bibitem{LY} 
E.~H.~Lieb, J.~Yngvason. Ground State Energy of the low density Bose Gas. {\it Phys. Rev. Lett.} {\bf 80} (1998), 2504–2507.    
\bibitem{NamRS}
P. T. Nam, N. Rougerie, R. Seiringer. Ground states of large bosonic systems: The Gross-Pitaevskii limit revisited. {\it Analysis and PDE}. {\bf 9} (2016), no. 2, 459–485.
\bibitem{NT}
 P. T. Nam, A. Triay. Bogoliubov excitation spectrum of trapped Bose gases in the Gross-Pitaevskii regime. Preprint arXiv:2106.11949.
\bibitem{NRS}
M. Napi{\'o}rkowski, R. Reuvers, J. P. Solovej. The Bogoliubov free energy functional I. Existence of minimizers and phase diagrams. {\it Arch. Ration. Mech. Anal.} {\bf 229}(3) (2018), 1037--1090.   
\bibitem{R}
D. W. Robinson. The Thermodynamic Pressure in Quantum Statistical Mechanics, {\it Lecture Notes in Physics} {\bf 9} (1971) 42–74.
	\bibitem{YY}
H.-T. Yau, J. Yin. The second order upper bound for the ground state energy of a Bose gas. {\it J. Stat. Phys.} {\bf 136}(3) (2009), 453--503.
\end{thebibliography}
\end{document}